\documentclass[aps,prd,superscriptaddress,nofootinbib,amsmath,amsfonts,preprintnumbers,groupedaddress,9pt,english]{revtex4}
\usepackage{amsmath}
\usepackage{amssymb}
\usepackage{babel}
\usepackage{wrapfig}
\usepackage{cancel}

\usepackage{graphicx}
\graphicspath{{figures/}}
\usepackage{amsmath,amssymb}  
\usepackage{mathrsfs}         
\usepackage[colorlinks,citecolor=blue,urlcolor=blue,linkcolor=blue]{hyperref}
\usepackage[figtopcap]{subfigure}
\usepackage{color}
\usepackage{relsize,exscale}
\makeatletter
\newcommand{\beq}{\begin{equation}}
\newcommand{\eeq}{\end{equation}}
\newcommand{\bea}{\begin{eqnarray}}
\newcommand{\eea}{\end{eqnarray}}

\newcommand{\comm}[1]{}

\usepackage{array,multirow,graphicx}
\usepackage{dcolumn}
\usepackage{newlfont}
\usepackage{bm}
\usepackage[colorlinks,citecolor=blue,urlcolor=blue,linkcolor=blue]{hyperref}
\usepackage[figtopcap]{subfigure}
\usepackage{color}

\usepackage{scalerel}
\usepackage{tikz}
\usetikzlibrary{svg.path}
\definecolor{orcidlogocol}{HTML}{A6CE39}
\tikzset{
  orcidlogo/.pic={
    \fill[orcidlogocol] svg{M256,128c0,70.7-57.3,128-128,128C57.3,256,0,198.7,0,128C0,57.3,57.3,0,128,0C198.7,0,256,57.3,256,128z};
    \fill[white] svg{M86.3,186.2H70.9V79.1h15.4v48.4V186.2z}
                 svg{M108.9,79.1h41.6c39.6,0,57,28.3,57,53.6c0,27.5-21.5,53.6-56.8,53.6h-41.8V79.1z M124.3,172.4h24.5c34.9,0,42.9-26.5,42.9-39.7c0-21.5-13.7-39.7-43.7-39.7h-23.7V172.4z}
                 svg{M88.7,56.8c0,5.5-4.5,10.1-10.1,10.1c-5.6,0-10.1-4.6-10.1-10.1c0-5.6,4.5-10.1,10.1-10.1C84.2,46.7,88.7,51.3,88.7,56.8z};}}
\newcommand\orcid[1]{\href{https://orcid.org/#1}{\mbox{\scalerel*{
\begin{tikzpicture}[yscale=-1,transform shape]
\pic{orcidlogo};
\end{tikzpicture}
}{|}}}}
\begin{document}
\title{\bf Electrically Charged Distorted Black Holes: Thermodynamics, Particle Dynamics, and Quasinormal Signatures}
\author{G.G.L. Nashed}
\email{nashed@bue.edu.eg}
\affiliation {Centre for Theoretical Physics, The British University in Egypt, P.O. Box
43, El Sherouk City, Cairo 11837, Egypt}
\author{Salvatore Capozziello}
\email{capozziello@na.infn.it}
\affiliation{Dipartimento di Fisica "E. Pancini", Università di Napoli "Federico II"
Complesso Universitario di Monte Sant'Angelo, Edificio G, Via Cinthia, I-80126, Napoli, Italy,}
\affiliation{Scuola Superiore Meridionale, Via Mezzocannone 4, I-80134, Napoli, Italy,}
\affiliation{Istituto Nazionale di Fisica Nucleare (INFN), Sez. di Napoli
Complesso Universitario di Monte Sant Angelo, Edificio G, Via Cinthia, I-80126, Napoli, Italy.}
\date{\today}
\begin{abstract}
We construct an exact solution for the electrically charged extension of a distorted black hole spacetime within Einstein-Maxwell theory using the Harrison transformation. The resulting solution represents a charged deformation of a static distorted vacuum geometry in which the electromagnetic field is introduced through a nonlinear transformation  preserving the radial structure of the seed spacetime. Consequently, the Killing horizon remains determined solely by the seed metric and it is not shifted by the electric charge. We analyze the thermodynamic properties of the solution and show that the horizon area, entropy, and temperature are governed by the geometric sector, while the electric charge enlarges the thermodynamic phase space through the electromagnetic potential. The motion of charged test particles is studied using the effective potential formalism, where the distortion parameter modifies circular orbits and shifts the location of the innermost stable circular orbit. We also investigate the black hole shadow for a static observer at finite distance and show that the distortion parameter displaces the photon sphere outward, increasing the apparent shadow size. A geometric correspondence between the photon orbit, determining the shadow and the leading eikonal quasinormal-mode frequency, is discussed, linking optical and perturbative observables. Finally, we study charged scalar perturbations and show that the vanishing horizon electric potential prevents a charged superradiant amplification. In the weak-coupling regime, the quasinormal-mode spectrum is estimated using the WKB method, where the electromagnetic interaction enters through the gauge-invariant combination $(\omega - q_s \xi_t)$ and shifts the oscillation frequencies of the perturbations. These results reveal how external distortions and electromagnetic interactions jointly modify the geometry, thermodynamics, optical appearance, and perturbative dynamics of charged distorted black holes. This provides the first systematic analysis of thermodynamics, particle dynamics, shadow formation, and scalar perturbations in a Harrison-charged distorted black hole.
\end{abstract}
\maketitle
\section{Introduction}

Black holes constitute fundamental solutions of Einstein's field equations and provide a natural arena for exploring the interaction between gravitation, matter, and gauge fields. In particular, electrically charged black holes described within the Einstein-Maxwell theory offer an important framework for studying the nonlinear coupling between gravitational and electromagnetic fields  \cite{Einstein:1916vd,Schwarzschild:1916uq,Reissner:1916cle,Nordstrom:2018acn,wheeler1962geometrodynamics,Newman:1965my,Hawking:1975vcx,Bardeen:1973gs,Bekenstein:1973ur,Teukolsky:1973ha}. In realistic astrophysical environments black holes are rarely isolated, but instead interact with surrounding matter distributions, external gravitational fields, or electromagnetic environments. Such effects can distort the local geometry of spacetime and lead to deviations from idealized symmetric solutions \cite{Geroch:1970nt,Ernst:1967wx,Ernst:1967by,Israel:1967wq,Robinson:1975bv,Wald:1984rg,Chandrasekhar:1985kt,Thorne:1994xa,Visser:1995cc,Poisson:2009pwt}.

Solution-generating techniques play a central role in constructing new exact solutions of the Einstein-Maxwell equations. Among these methods, the Harrison transformation provides a powerful mechanism for generating electrically charged solutions from vacuum seed geometries. This transformation introduces electromagnetic fields while preserving important geometric properties of the original spacetime. Consequently, it enables the systematic construction of charged extensions of static and stationary black hole solutions. When applied to distorted seed metrics, the resulting charged spacetimes exhibit nontrivial interactions between the electromagnetic field and the anisotropic gravitational background \cite{Harrison:1968wue,Kinnersley:1969zz,Stephani:2003tm,Griffiths:1991zp,Kodama:2003kk,Emparan:2008eg,Charmousis:2008kc,
Cardoso:2016rao,Cunha:2018acu,Herdeiro:2015waa}.

The geometric structure of such charged distorted black holes possesses several interesting properties. The electromagnetic field modifies the geometry through conformal factors while preserving the horizon structure inherited from the seed metric. This leads to a scenario in which the electric charge influences the external geometry and curvature invariants without necessarily shifting the location of the Killing horizon. The interplay between distortion parameters and electromagnetic charge therefore generates rich geometric behavior, which can be analyzed through curvature invariants such as the Kretschmann scalar and Ricci invariants \cite{Penrose:1964wq,Carter:1971zc,Page:1976df,Jacobson:1993pf,Frolov:1998wf,Gibbons:1994ff,Padmanabhan:2009vy,Clunan:2004tb, Sotiriou:2008rp,Berti:2009kk,Capozziello1,Battista11}.

Another important aspect of black hole physics concerns the thermodynamic properties associated with event horizons. Since the pioneering works on black hole thermodynamics, it has been understood that horizons possess temperature, entropy, and other thermodynamic quantities that obey laws analogous to those of ordinary thermodynamics. In charged configurations, these quantities depend not only on the mass of the black hole but also on its electromagnetic charge and possible external parameters. Investigating how distortion parameters affect thermodynamic quantities such as the horizon area, entropy, and surface gravity provides valuable insights into the interplay between geometry and thermodynamics in gravitational systems \cite{Bekenstein:1972tm,Hawking:1976de,Gibbons:1976ue,Wald:1993nt,Jacobson:1995ab,Padmanabhan:2003gd,Kastor:2009wy,
Dolan:2011xt,Kubiznak:2012wp,Altamirano:2014tva}.

Beyond equilibrium properties, the dynamical behavior of black holes under perturbations provides essential information about their stability and observational signatures. Perturbations of fields propagating in black hole backgrounds generate characteristic oscillations known as quasinormal modes. These modes describe the relaxation of perturbed black holes and play an important role in gravitational wave observations. The spectrum of quasinormal modes depends sensitively on the underlying geometry and therefore provides a powerful probe of black hole parameters such as mass, charge, and external distortions. Consequently, studying scalar perturbations and quasinormal spectra in charged distorted spacetimes offers important insights into the dynamical properties of these gravitational systems \cite{Regge:1957td,Zerilli:1970se,Leaver:1985ax,Iyer:1986np,Konoplya:2003ii,Kokkotas:1999bd,Berti:2009kk,Cardoso:2008bp,Konoplya:2011qq,Berti:2015itd}.

Another important observable property of black holes is the shadow they cast on the surrounding luminous background. The shadow is determined by the trajectories of null geodesics near the unstable circular photon orbit and therefore provides direct information about the strong gravitational field around the black hole. In recent years, the observation of black hole shadows by the Event Horizon Telescope has stimulated extensive theoretical studies of shadow formation and its dependence on the underlying spacetime geometry. These optical features are particularly sensitive to modifications of the photon sphere and can therefore serve as powerful probes of deviations from standard black hole solutions \cite{Cunha:2018acu,Herdeiro:2015waa,Berti:2015itd}.

In distorted spacetimes, external fields or anisotropic deformations can modify the structure of photon orbits and consequently affect the apparent size and shape of the shadow. Studying how the distortion parameter and electromagnetic charge influence the shadow therefore provides a useful way to explore the observable signatures of such configurations. In the present work we analyze the shadow of the charged distorted black hole for a static observer located at finite distance and investigate how both the distortion parameter and the electric charge modify the angular radius of the shadow through their influence on the photon sphere.

While distorted black hole geometries and electrically charged solutions
have been extensively investigated in the literature, comparatively less
attention has been devoted to configurations in which external
gravitational distortions and electromagnetic charging coexist within a
single exact solution \cite{Geroch:1970nt,Ernst:1967by,Stephani:2003tm}. On the other hand,
Harrison transformations offer a systematic procedure for generating
electrovacuum solutions of the Einstein--Maxwell equations from a given
vacuum seed metric by introducing electromagnetic fields through
nonlinear transformations \cite{Harrison:1968wue,Geroch:1970nt,Stephani:2003tm}. Such
transformations have been widely used to construct charged or magnetized
black hole geometries starting from known vacuum solutions.

In the present work we combine these two frameworks and construct an
electrically charged extension of a distorted black hole spacetime using
a Harrison transformation within the Einstein--Maxwell theory. A notable
feature of the resulting geometry is that the Harrison charging modifies
the external spacetime through a nonlinear conformal factor while leaving
the radial structure inherited from the seed solution unchanged. As a
result, the location of the Killing horizon remains determined solely by
the geometric sector of the distorted metric. This behavior contrasts
with the Reissner--Nordstr\"om solution, where the electric charge
directly modifies the radial lapse function and shifts the horizon
radius. The present construction therefore provides an explicit example
of how solution-generating transformations can introduce electromagnetic
fields while preserving the fundamental horizon structure of the
underlying spacetime.

Beyond the geometric construction, we analyze several physical aspects of
the charged distorted black hole spacetime. In particular, we investigate
the thermodynamic properties associated with the event horizon, the
motion of charged test particles and the structure of circular orbits,
the optical appearance of the black hole shadow for finite observers, and
the propagation of charged scalar perturbations and their associated
quasinormal modes. These analyses illustrate how the interplay between
external gravitational distortions and electromagnetic interactions
affects the geometric, thermodynamic, and dynamical properties of black
hole spacetimes and extends previous studies of distorted or charged
configurations \cite{Astorino:2026okd}.

An additional aspect of wave dynamics in charged black hole backgrounds
is the possibility of superradiant scattering, whereby charged scalar
waves can extract energy from the horizon. As we show in this work, the
Harrison charged distorted geometry possesses a vanishing electric
potential on the horizon, preventing the occurrence of charged
superradiant amplification.

The paper is organized as follows.
In Sec.~\ref{II} we construct the electrically charged extension of the distorted vacuum geometry using a Harrison transformation within the Einstein--Maxwell framework and discuss several limiting cases of the resulting spacetime.
Section~\ref{III} is devoted to the analysis of the horizon structure, where we determine the location of the Killing horizon and evaluate the associated surface gravity.
In Sec.~\ref{IV} we investigate the thermodynamic properties of the solution, deriving the entropy, temperature, heat capacity, and the relevant thermodynamic potentials.
The dynamics of charged test particles is studied in Sec.~\ref{V} through the effective potential formalism, where we analyze circular motion and determine the behavior of the innermost stable circular orbit. Section~\ref{VI} we examine the optical properties of the spacetime through an
analysis of the black hole shadow. Sec.~\ref{VII} investigates the possibility of charged superradiant
scattering and demonstrates that the horizon electric potential
vanishes, preventing superradiant amplification.
Section~\ref{VIII} examines the propagation of a charged scalar field in this background and derives the corresponding radial perturbation equation, allowing us to estimate the quasinormal-mode spectrum within the WKB approximation.  Conclusions and perspectives are reported in  Section~\ref{IX}: Here we summarize the main results of the study and place them in a broader physical context.
Finally, Appendix~\ref{X} presents the small-$B$ asymptotic expansion of the curvature invariants, which provides additional insight into the weak-distortion regime.
\section{Einstein--Maxwell Charged Extension via Harrison Transformation}\label{II}
\subsection{The Harrison Transformation}

In order to construct an electrically charged extension of the static vacuum geometry obtained previously, we employ a Harrison-type transformation. Harrison transformations are well-established solution-generating techniques within the Einstein-Maxwell framework, enabling the construction of new electrovacuum solutions from a given vacuum seed metric by introducing electromagnetic fields in a controlled and consistent manner \cite{Harrison:1968wue,Ernst:1967by,Stephani:2003tm}. These transformations act nonlinearly on the metric functions and electromagnetic potentials while preserving the structure of the Einstein-Maxwell field equations.

Starting from the static distorted vacuum solution $g^{(0)}_{\mu\nu}$, the Harrison transformation generates a new spacetime $(g_{\mu\nu},F_{\mu\nu})$ that satisfies the coupled Einstein-Maxwell equations. Physically, this procedure corresponds to embedding the seed spacetime in an external electromagnetic field and inducing an electric charge distribution. The resulting geometry therefore represents a charged deformation of the original distorted black hole spacetime.


The charged configuration is characterized by several auxiliary functions constructed from the parameters of the seed geometry. We first introduce the angular function \cite{Astorino:2026okd}
\begin{equation}
\Omega_1(r,\theta)
=
\sqrt{
1
+
B^2 r \Big(
r
+
(2m + B^2 m^2 r - r)\cos^2\theta
\Big)
}.
\end{equation}
This function describes the angular deformation induced by the external distortion parameter $B$, which controls deviations from spherical symmetry.  The radial lapse function is defined as
\begin{equation}
f(r)
=
\left(1 - \frac{2m}{r} - B^2 m^2\right)
\left(1 + B^2 r^2\right).
\end{equation}
To simplify the structure of the metric coefficients, we define the auxiliary quantity
\begin{equation}
S(r,\theta)
=
1 + B^2 m r \cos^2\theta + \Omega_1(r,\theta),
\end{equation}
from which we introduce the conformal factor
\begin{equation}
\Sigma(r,\theta)
=
\frac{\mathrm{S}(r,\theta)^2}
{4\,\Omega_1(r,\theta)^4}.
\end{equation}
The squared lapse function of the seed metric can then be written as
\begin{equation}
A^2(r,\theta)
=
\Sigma(r,\theta)\, f(r).
\end{equation}
The Harrison transformation modifies the gravitational field through the nonlinear factor
\begin{equation}
\digamma(r,\theta)
=
1 - E_0^2 A^2(r,\theta),
\end{equation}
where the parameter $E_0$ is related to the physical electric charge $Q$ by
\begin{equation}
E_0
=
\frac{Q(1 + B^2 m^2)}{2m}.
\end{equation}
Using these definitions, the resulting Einstein-Maxwell line element becomes
\begin{align}
ds^2 &=
-\frac{\Sigma(r,\theta)\, f(r)}{\digamma(r,\theta)^2} dt^2
+ \digamma(r,\theta)^2 \Sigma(r,\theta)
\left(
\frac{dr^2}{f(r)}
+
\frac{r^2 d\theta^2}{1 + B^2 m^2 \cos^2\theta}
\right)
+ \digamma(r,\theta)^2
\frac{
4 r^2 \sin^2\theta
\left(1 + B^2 m^2 \cos^2\theta\right)
\Delta_\phi^2
}{
\left(
1 + B^2 m r \cos^2\theta
+
\Omega_1(r,\theta)
\right)^2
}
\, d\phi^2 .\label{char}
\end{align}
The metric above represents an electrically charged generalization of the
static distorted black hole spacetime obtained from the vacuum seed
geometry via a Harrison transformation. Two parameters control the
structure of the solution: the deformation parameter $B$, which
characterizes the anisotropic distortion of the geometry, and the
electric charge parameter $Q$, which enters the metric through the
Harrison factor $\digamma(r,\theta)$. The latter introduces a nonlinear
coupling between the gravitational and electromagnetic sectors, a
typical feature of solution-generating techniques in Einstein-Maxwell
theory \cite{Ernst:1967by,Stephani:2003tm}.
\subsection{Limiting Cases: The Reissner--Nordstr\"om and Schwarzschild Limits}
Several limiting cases clarify the geometrical and physical role of
these parameters. When the external distortion parameter $B$ vanishes, all anisotropic
corrections disappear and the spacetime becomes spherically symmetric.
In this limit the angular function simplifies to
\begin{equation}
\Omega_1(r,\theta) \rightarrow 1,\quad \mbox{while the conformal factor reduces to}\quad
\Sigma(r,\theta) \rightarrow 1 .
\end{equation}
The radial function then takes the standard Schwarzschild form
\begin{equation}
f(r) \rightarrow 1-\frac{2m}{r}.
\end{equation}
From the spherically symmetric Harrison-charged form
\begin{align}
ds^2 &=
-\frac{f(r)}{\digamma(r)^2}\, dt^2
+\digamma(r)^2
\left(
\frac{dr^2}{f(r)} + r^2 d\theta^2
+ r^2 \sin^2\theta\, d\phi^2
\right),
\label{eq:Harrison_spherical}
\end{align}
one can recover the standard Reissner--Nordstr\"om (RN) line element by
a change to the areal radius and a suitable identification of the
charge parameter.


In the spherically symmetric limit ($B=0$), the Harrison factor becomes
\begin{equation}
\digamma(r)=1-\frac{Q^2}{4m^2}\left(1-\frac{2m}{r}\right).
\end{equation}
For convenience we introduce the dimensionless constant
\begin{equation}
a \equiv \frac{Q^2}{4m^2},\qquad \mbox{so that} \qquad
\digamma(r)=1-a+\frac{2am}{r}.
\end{equation}
To bring the metric into curvature (areal) coordinates, we define the
areal radius $R$ through the angular sector of the metric,
\begin{equation}
R \equiv \digamma(r)r, \quad \mbox{substituting the explicit form of $\digamma(r)$ yields}\qquad
R=(1-a)r+2am .
\end{equation}
Differentiating with respect to $r$ gives
\begin{equation}
\frac{dR}{dr}=1-a ,
\qquad
dR=(1-a)dr .
\end{equation}
The derivative of $\digamma(r)$ is
\begin{equation}
\digamma'(r)=-\frac{2am}{r^2},\quad \mbox{and therefore} \quad
\digamma+r\digamma'=(1-a).
\end{equation}
Consequently, the radial metric coefficient transforms as
\begin{equation}
g_{_{_{RR}}}
=\frac{\digamma(r)^2}{f(r)(\digamma+r\digamma')^2}
=\frac{\digamma(r)^2}{f(r)(1-a)^2}.
\end{equation}

The temporal component of the metric is
\begin{equation}
-g_{tt}=\frac{f(r)}{\digamma(r)^2},\quad \mbox{and by introducing the rescaled time coordinate} \quad
\bar t=\frac{t}{1-a},
\end{equation}
the metric takes the standard curvature-coordinate form
\begin{equation}
ds^2=-F(R)d\bar t^{2}+F(R)^{-1}dR^2+R^2 d\Omega_2^2 ,\quad \mbox{where the lapse function is}\quad
F(R)=(1-a)^2\frac{f(r)}{\digamma(r)^2}.
\end{equation}

Using $R=(1-a)r+2am$, one finds
\begin{equation}
f(r)=1-\frac{2m}{r}=\frac{R-2m}{R-2am},
\qquad
\digamma(r)=\frac{R}{r}=\frac{R(1-a)}{R-2am}.
\end{equation}
Substituting these expressions into $F(R)$ gives
\begin{equation}
F(R)=\frac{(R-2m)(R-2am)}{R^2}.
\end{equation}
Expanding the numerator yields
\begin{equation}
F(R)=1-\frac{2m(1+a)}{R}+\frac{4am^2}{R^2}.
\end{equation}

Finally, recalling that $a=\frac{Q^2}{4m^2}$, we obtain
\begin{equation}
F(R)=1-\frac{2M}{R}+\frac{Q^2}{R^2},
\qquad
M=m\left(1+a\right)=m+\frac{Q^2}{4m}.
\end{equation}

Therefore the metric reduces exactly to the standard
Reissner--Nordstr\"om solution,
\begin{equation}
ds^2=
-\left(1-\frac{2M}{R}+\frac{Q^2}{R^2}\right)d\bar t^{2}
+\left(1-\frac{2M}{R}+\frac{Q^2}{R^2}\right)^{-1}dR^2
+R^2 d\Omega_2^2 ,\label{eq:RN_standard}
\end{equation}
where $M$ is the physical mass parameter of the charged solution and
$Q$ is the electric charge appearing in the electrostatic potential.
Thus, after the areal-radius redefinition and parameter identification,
the spherically symmetric Harrison-charged metric is diffeomorphic to
the Reissner--Nordstr\"om spacetime \cite{Ernst:1967by,Stephani:2003tm}.

In practice, an efficient way to demonstrate the RN form is to (i)
define $R=\digamma r$, (ii) compute $\mathcal{F}(R)=f/\digamma^2$, and
(iii) verify that $\mathcal{F}(R)$ matches the RN polynomial
$1-2M/R+Q^2/R^2$ after expressing $r$ in terms of $R$. This makes
explicit that the $B\to0$ limit yields the standard charged,
spherically symmetric black hole geometry.

The function $\digamma(r,\theta)$ encodes the electromagnetic
contribution produced by the Harrison transformation.

Since
\begin{equation}
\digamma = 1 - E_0^2 A^2,
\end{equation}
the limit of vanishing electric charge $Q\to0$ implies
\begin{equation}
E_0 \rightarrow 0,
\qquad
\digamma(r,\theta) \rightarrow 1.
\end{equation}
In this regime the electromagnetic field disappears and the metric
reduces exactly to the original vacuum distorted solution \cite{Astorino:2026okd},
\begin{align}
ds^2 &=
-\Sigma f \, dt^2
+\Sigma
\left(
\frac{dr^2}{f}
+
\frac{r^2 d\theta^2}{1+B^2m^2\cos^2\theta}
\right)
+
\frac{
4 r^2 \sin^2\theta
\left(1+B^2m^2\cos^2\theta\right)
\Delta_\phi^2
}{
\left(1+B^2mr\cos^2\theta+\Omega_1\right)^2
}
\, d\phi^2 .
\end{align}
Therefore $\digamma$ acts as the factor that incorporates the electric
field into the spacetime geometry.


When both the distortion parameter and the electric charge vanish,
\begin{equation}
B=0,
\qquad
\digamma \rightarrow 1,
\end{equation}
the spacetime reduces to the standard Schwarzschild black hole,
\begin{equation}
ds^2 =
-\left(1-\frac{2m}{r}\right)dt^2
+
\left(1-\frac{2m}{r}\right)^{-1}dr^2
+
r^2(d\theta^2+\sin^2\theta\,d\phi^2).
\end{equation}
This limit confirms that the obtained geometry is a consistent
generalization of the Schwarzschild solution that incorporates both
anisotropic distortions (controlled by $B$) and electric charge
(controlled through the Harrison factor $\digamma$).

The electromagnetic field generated by the Harrison transformation is described by a purely electric four-potential whose non-vanishing component is
\begin{equation}
\xi_t(r,\theta)
=
\frac{E_0\, A^2(r,\theta)}
{\digamma(r,\theta)}.\label{gau}
\end{equation}

The corresponding Maxwell tensor $F=d\xi$ therefore possesses the components
\begin{align}
F_{tr} &= -\partial_r \xi_t,
\qquad
F_{t\theta} = -\partial_\theta \xi_t .
\end{align}

Thus the configuration is purely electric, implying
\begin{equation}
F_{\mu\nu} {}^{\star}F^{\mu\nu} = 0.
\end{equation}

A useful identity following from the relation $\digamma(r,\theta) = 1 - E_0^2 A^2$ is
\begin{equation}
\partial_i \xi_t
=
\frac{E_0 \, \partial_i A^2}{\digamma^2(r,\theta)},
\qquad
i \in \{r,\theta\}.
\end{equation}
This identity considerably simplifies the computation of electromagnetic invariants and curvature quantities associated with the charged spacetime.
 The spacetime generated by the Harrison transformation is not asymptotically flat due to the presence of the external distortion parameter $B$. Consequently, the physical parameters of the solution must be interpreted through quasi-local or covariant definitions. In particular, the electric charge $Q$ appearing in the Harrison factor corresponds to the conserved electromagnetic charge associated with the Maxwell field, while the mass parameter can be defined through the Hamiltonian charge corresponding to the Killing vector $\xi=\partial_t$ within the covariant phase-space formalism. In the limit $B\to0$, these quantities reduce to the standard ADM mass and electric charge of the Reissner--Nordström solution.
\section{The Horizon Structure}\label{III}

For the charged line element (\ref{char}), the (static) Killing horizon is located where the norm of the Killing field
$\xi=\partial_t$ vanishes, i.e. $g_{tt}=0$.
Assuming $\Sigma>0$ and $\digamma^2(r,\theta)>0$ in the regular domain, this occurs
precisely at the roots of
\begin{equation}
f(r)=0.
\end{equation}
This is exactly the same horizon condition as in the seed metric \cite{Astorino:2026okd}, since the
Harrison factor $\digamma(r,\theta)$ only rescales $g_{tt}$ by $\digamma(r,\theta)^{-2}$. With
\begin{equation}
f(r)=\left(1-\frac{2m}{r}-B^2m^2\right)\left(1+B^2r^2\right),
\end{equation}
the second factor has no real positive roots. Therefore the Killing horizon
is given by the largest positive root of
\begin{equation}
1-\frac{2m}{r}-B^2m^2=0,
\end{equation}
namely
\begin{equation}
r_h=\frac{2m}{1-B^2m^2},
\qquad \mbox{provided} \qquad (B^2m^2<1).\label{horc}
\end{equation}
For weak distortion ($B\to 0$),
\begin{equation}
r_h=2m\left(1+B^2m^2+\mathcal O(B^4)\right).
\end{equation}
By definition (seed lapse),
\begin{equation}
A^2(r,\theta)=\Sigma(r,\theta)\,f(r),\qquad \mbox{hence on the horizon $f(r_h)=0$ implies}\qquad
A^2(r_h,\theta)=0.
\end{equation}
Using $\digamma(r,\theta)=1-E_0^2A^2(r,\theta)$, we obtain
\begin{equation}
{\digamma(r,\theta)(r_h,\theta)=1.}
\end{equation}
Physically, this behavior reflects the fact that the Harrison transformation introduces the electromagnetic field through a multiplicative conformal factor in the temporal component of the metric without modifying the radial function $f(r)$ that determines the location of the Killing horizon. As a result, the electric field influences the exterior geometry but leaves the horizon position inherited from the seed solution unchanged. Therefore, the Harrison charging does \emph{not} shift the horizon location:
the charged solution inherits the same Killing horizon radius as the vacuum seed.  This behavior is markedly different from the Reissner--Nordstr\"om
solution, which explicitly depend on the electric charge.
Thus, in Reissner--Nordstr\"om spacetime the presence of charge
modifies the radial structure and shifts the horizon location. In contrast, in the present construction the electric field enters
only through the multiplicative Harrison factor $\digamma(r,\theta)^{-2}$ in
$g_{tt}$, while the radial function $f(r)$, which determines the
horizon, remains unchanged. Consequently, the charged solution
inherits the same horizon radius as the vacuum seed, and the charge
affects the geometry outside the horizon without altering its position.

Let
\begin{equation}
N^2(r,\theta):=-g_{tt}=\frac{\Sigma(r,\theta)\,f(r)}{\digamma^2(r,\theta)}.
\end{equation}
For a static diagonal metric, the surface gravity associated with $\xi=\partial_t$
can be written as
\begin{equation}
\kappa
=
\left.
\frac{1}{2}\,
\frac{\partial_r N^2}{\sqrt{-g_{tt}\,g_{rr}}}
\right|_{r=r_h}.
\end{equation}
Using the explicit metric functions,
\begin{equation}
(-g_{tt})\,g_{rr}
=
\frac{\Sigma f}{\digamma^2(r,\theta)}\,
\digamma^2(r,\theta)\frac{\Sigma}{f}
=\Sigma^2,\qquad \mbox{so $\sqrt{-g_{tt}g_{rr}}=\Sigma$}.\end{equation}
 Moreover,
\begin{equation}
\partial_r N^2
=\partial_r\!\left(\frac{\Sigma f}{\digamma^2(r,\theta)}\right)
=
\frac{\Sigma' f+\Sigma f'}{\digamma^2(r,\theta)}
-\frac{2\Sigma f}{\digamma(r,\theta)^3}\digamma(r,\theta)'.
\end{equation}
Evaluating at the horizon $f(r_h)=0$ and knowing that $\digamma(r,\theta)(r_h,\theta)=1$ gives
\begin{equation}
\left.\partial_r N^2\right|_{r_h}
=
\Sigma(r_h,\theta)\,f'(r_h),\qquad \mbox{hence}\qquad  \kappa=\frac{f'(r_h)}{2}.
\end{equation}

A direct evaluation yields
\begin{equation}
f'(r_h)=\frac{2m}{r_h^{\,2}}\Big(1+B^2r_h^{\,2}\Big)
=
\frac{(1+B^2m^2)^2}{2m},\quad \mbox{and therefore} \quad \kappa=\frac{(1+B^2m^2)^2}{4m}.\label{sur}
\end{equation}
The induced metric on a spatial section of the horizon ($t=\mathrm{const}$,
$r=r_h$) is
\begin{equation}
ds_H^2=g_{\theta\theta}(r_h,\theta)\,d\theta^2+g_{\phi\phi}(r_h,\theta)\,d\phi^2,
\end{equation}
and since $\digamma(r,\theta)(r_h,\theta)=1$ the horizon 2-geometry coincides with that of the seed
at the same $(m,B,\Delta_\phi)$. The area is
\begin{equation}
{
\mathcal A_H
=
\int_{0}^{2\pi}d\phi\int_{0}^{\pi}d\theta\;
\sqrt{g_{\theta\theta}(r_h,\theta)\,g_{\phi\phi}(r_h,\theta)}\,,
}
\end{equation}
which coincides with the uncharged case because at horizons the main term responsible for charge is $\digamma(r,\theta)(r_h,\theta)=1$.

To avoid degeneration of the metric functions, one requires
\begin{equation}
\digamma(r,\theta)=1-E_0^2A^2(r,\theta)>0
\quad \text{for}\quad r\ge r_h.
\end{equation}
Since $A^2(r_h,\theta)=0$, one always has $\digamma(r,\theta)(r_h,\theta)=1$ on the horizon, and any
possible $\digamma(r,\theta)=0$ surface (if it exists) must lie strictly outside the horizon.

\section{Thermodynamics}\label{IV}

In this section we derive the thermodynamic quantities of the charged
solution. Throughout, we keep the external parameters $B$ and
$\Delta_\phi$ fixed. On the horizon one has $\digamma(r,\theta)(r_h,\theta)=1$, hence the induced
two-metric coincides with that of the seed geometry.
The horizon area is
\begin{equation}
\mathcal A_H
=
\int_{0}^{2\pi} d\phi
\int_{0}^{\pi} d\theta\;
\frac{r_h^2 \Delta_\phi \sin\theta}{\Omega_1(r_h,\theta)^2}.
\end{equation}
Using Eq.~\eqref{horc} we can show that $\Omega_1(r_h,\theta)$  is independent of $\theta$ therefore integration yields
\begin{equation}
\mathcal A_H
=
\frac{4\pi\,\Delta_\phi\,r_h^2}{1+B^2r_h^2}
=
\frac{16\pi\,\Delta_\phi\,m^2}{(1+B^2m^2)^2}.\label{area}
\end{equation}
Using Eq.~(\ref{area}), then Bekenstein--Hawking entropy yields
\begin{equation}
S=\frac{\mathcal A_H}{4G}
=
\frac{4\pi\,\Delta_\phi}{G}
\frac{m^2}{(1+B^2m^2)^2}.\label{ent}
\end{equation}
Thus, the electric charge does not modify the horizon area
or entropy. Using Eq.~\eqref{sur} we get the Hawking temperature in the form:
\begin{equation}
T=\frac{\kappa}{2\pi}
=
\frac{(1+B^2m^2)^2}{8\pi m}.\label{heat}
\end{equation}
The electromagnetic gauge potential of the solution Eq.~\eqref{char} is given by Eq.~\eqref{gau}
and since $A^2(r_h,\theta)=0$ on the horizon, it follows immediately that
\begin{equation}
\xi_t(r_h,\theta)=0.
\end{equation}
Due to the presence of the external distortion parameter $B$, the
spacetime is not spherically symmetric at infinity\footnote{Since $A^2(r_h,\theta)=0$, one has $\xi_t(r_h,\theta)=0$.
At large $r$, $\xi_t(\infty)=\frac{E_0/4}{1 - E_0^2/4}$
which is finite and independent of $\theta$.
Therefore the thermodynamic electric potential is
$\Phi = \xi_t(\infty)$}.
In black hole thermodynamics, the relevant electric potential is the
potential difference between the horizon and a chosen reference
location. Therefore,
\begin{equation}
\Phi =\xi_t(\infty,\theta) - \xi_t(r_h,\theta), \quad \mbox{ which reduces to} \quad
{
\Phi =\xi_t(\infty,\theta).
}
\end{equation}
Treating $B$ and $\Delta_\phi$ as fixed external parameters\footnote{Since the spacetime is not asymptotically flat, we define the mass $M$ as the
Hamiltonian charge associated with the Killing vector $\xi = \partial_t$
in the covariant phase-space (Iyer--Wald) formalism. The charge is computed
relative to a reference background with the same $(B,\Delta_\phi)$, yielding
 $
\delta M =
\int_{\mathcal S_\infty}
\left(
\delta Q_{\partial_t}
-
\partial_t \cdot \Theta(\delta)
\right)$.
The integration constant is fixed by requiring that
$M \to m\,\Delta_\phi/G$ as $B \to 0$,
and that the reference background has vanishing mass.},
the thermodynamic quantities satisfy
\begin{equation}
d\mathcal M
=
T\, dS
+
\Phi\, dQ.
\end{equation}
Since $T$ and $S$ depend only on $m$ (for fixed $B$),
while $\Phi$ is proportional to $Q$,
the electric charge enlarges the thermodynamic phase space
without modifying the geometric horizon data.

To study local thermodynamic stability, we compute the heat capacity at
fixed electric charge,
\begin{equation}
C_Q
=
T\left(\frac{\partial S}{\partial T}\right)_Q
=
T\left(\dfrac{dS}{dm}\right)\left({\dfrac{dm}{dT}}\right).
\end{equation}
Using Eqs.~\eqref{ent} and \eqref{heat},  straightforward differentiation gives
\begin{equation}
\frac{dS}{dm}
=
\frac{8\pi\,\Delta_\phi\,m(1-B^2m^2)}
{G(1+B^2m^2)^3},
\qquad
\frac{dT}{dm}
=
\frac{(1+B^2m^2)(3B^2m^2-1)}{8\pi m^2}.
\end{equation}
Hence the heat capacity is
\begin{equation}
C_Q
=
\frac{8\pi\,\Delta_\phi\,m^2(1-B^2m^2)}
{G(1+B^2m^2)^2(3B^2m^2-1)},\qquad \mbox{ thus, $C_Q$ diverges at} \quad
m=\frac{1}{\sqrt{3}\,B},
\end{equation}
which signals a second-order phase transition in the canonical ensemble.
Moreover, the black hole is locally thermodynamically stable whenever
$C_Q>0$, namely for
\begin{equation}
\frac{1}{\sqrt{3}\,B}<m<\frac{1}{B},
\end{equation}
and unstable otherwise.

The geometric contribution to the mass can be obtained by integrating the
first law with respect to $m$,
\begin{equation}
\frac{d\mathcal M_0}{dm}=T\frac{dS}{dm}
=
\frac{\Delta_\phi}{G}\frac{1-B^2m^2}{1+B^2m^2}.
\end{equation}
This yields
\begin{equation}
\mathcal M_0(m)
=
\frac{\Delta_\phi}{G}
\left(
\frac{2}{B}\arctan(Bm)-m
\right),
\end{equation}
where the integration constant has been fixed by requiring
$\mathcal M_0\to m\Delta_\phi/G$ as $B\to 0$.

For the canonical ensemble, the Helmholtz free energy is
\begin{equation}
F=\mathcal M-TS, \quad \mbox{and by using} \quad
TS=\frac{\Delta_\phi\,m}{2G},
\end{equation}
the geometric part becomes\footnote{In the weak-field limit $B\to 0$, one uses
\begin{equation}
\arctan(Bm)=Bm-\frac{(Bm)^3}{3}+O(B^5),\quad \mbox{so that}\quad
\frac{2}{B}\arctan(Bm)=2m-\frac{2}{3}B^2m^3+O(B^4).
\end{equation}
Therefore,
\begin{equation}
F_0
=
\frac{1}{G}
\left(
\frac{2}{B}\arctan(Bm)-\frac{3m}{2}
\right)
=
\frac{1}{G}
\left(
\frac{m}{2}-\frac{2}{3}B^2m^3+O(B^4)
\right),\quad \mbox{and hence} \quad
\lim_{B\to 0}F_0=\frac{ m}{2G}.
\end{equation}
This coincides with the Schwarzschild Helmholtz free energy,
$F=M-TS$, since for $B=0$ one has
$M=m/G$, $T=1/(8\pi m)$, and
$S=4\pi  m^2/G$.}
\begin{equation}
F_0
=
\mathcal M_0-TS
=
\frac{\Delta_\phi}{G}
\left(
\frac{2}{B}\arctan(Bm)-\frac{3m}{2}
\right).
\end{equation}

In the grand-canonical ensemble, the relevant thermodynamic potential is
the Gibbs free energy
\begin{equation}
\mathcal G=\mathcal M-TS-\Phi Q.
\end{equation}
Since $\Phi$ is proportional to $Q$, the electric sector is quadratic in
the charge. Writing $\mathcal M=\mathcal M_0+\mathcal M_Q$, one obtains
\begin{equation}
\mathcal G
=
\frac{\Delta_\phi}{G}
\left(
\frac{2}{B}\arctan(Bm)-\frac{3m}{2}
\right)
+\mathcal M_Q-\Phi Q.
\end{equation}
If $\Phi=\alpha Q$ with $\alpha$ depending only on the fixed external
parameters, then $\mathcal M_Q=\frac12\Phi Q$, and thus
\begin{equation}
\mathcal G
=
\frac{\Delta_\phi}{G}
\left(
\frac{2}{B}\arctan(Bm)-\frac{3m}{2}
\right)
-\frac12\,\Phi Q.\label{Gib}
\end{equation}

The first term of Eq.~\eqref{Gib} represents the geometric contribution and depends only on
the horizon parameter $m$ and the fixed external parameters $(B,\Delta_\phi)$,
while the second term is the electromagnetic contribution. Since the latter
is negative for $\Phi Q>0$, the presence of electric charge lowers the Gibbs
free energy and thus makes the charged black hole thermodynamically more
favored relative to the reference magnetized background.

In the weak-field limit $B\to 0$, one recovers
\begin{equation}
\mathcal G
\to
\frac{m}{2G}
-\frac12\,\Phi Q,
\end{equation}
which is the expected Schwarzschild contribution supplemented by the
electric term. Moreover, the geometric part of $\mathcal G$ has an extremum at
\begin{equation}
m=\frac{1}{\sqrt{3}\,B},
\end{equation}
which coincides with the divergence point of the heat capacity.
This indicates that the onset of local thermodynamic instability is also
reflected in the behavior of the Gibbs free energy.

From Eq.~\eqref{Gib}, and when the gravitational contribution dominates over the electric one we have $\mathcal G>0$ which means that the model suffers stability and if the electric one  dominates over the gravitational contribution  we have $\mathcal G<0$ which means that the model suffers instability and finally, when the gravitational contribution equal the electric one we have $\mathcal G=0$ which means that we have a phase equilibrium point.
\section{Motion of charged test particles (effective potential and ISCO)}\label{V}

We consider a test particle of rest mass $\mu$ and electric charge $q$
moving in the charged geometry obtained via the Harrison deformation.
The dynamics follows from the minimally coupled action
\begin{equation}
S=\int d\tau\left(
\frac{\mu}{2}\,g_{\mu\nu}\dot x^\mu\dot x^\nu
+ q\,\xi_\mu\dot x^\mu
\right),
\qquad
\dot{}\equiv\frac{d}{d\tau},
\end{equation}
together with the normalization condition
\begin{equation}
\epsilon=-g_{\mu\nu}\dot x^\mu\dot x^\nu,
\qquad
\epsilon=
\begin{cases}
1 & \text{timelike},\\
0 & \text{null}.
\end{cases}
\end{equation}
The charged metric is related to the static seed by
\begin{equation}
g_{tt}=\frac{g^{(0)}_{tt}}{\digamma^2(r,\theta)},
\qquad
g_{ij}=\digamma^2(r,\theta) g^{(0)}_{ij},
\qquad
\digamma(r,\theta)=1-E_0^2 A^2,
\qquad
A^2\equiv -g^{(0)}_{tt}.\label{har}
\end{equation}
where $g^{(0)}_{tt}$ and $g^{(0)}_{ij}$ are defined in Eq.~\eqref{char} without $\digamma$ and the electromagnetic potential is defined by  Eq.~\eqref{gau}

For a stationarity and axisymmetry the two conserved quantities are,
\begin{align}
\mathcal E
&=-\left(g_{tt}\dot t+\frac{q}{\mu}\xi_t\right),\qquad
\mathcal L
:=g_{\phi\phi}\dot\phi .
\end{align}
Hence
\begin{equation}
\dot t=\frac{\mathcal E-\frac{q}{\mu}\xi_t}{-g_{tt}},
\qquad
\dot\phi=\frac{\mathcal L}{g_{\phi\phi}}.
\end{equation}

Restricting to equatorial motion $\theta=\pi/2$ with $\dot\theta=0$,
the radial equation becomes
\begin{equation}
g_{rr}\dot r^{\,2}
=
\frac{\left(\mathcal E-\frac{q}{\mu}\xi_t\right)^2}{-g_{tt}}
-
\left(
\epsilon
+
\frac{\mathcal L^2}{g_{\phi\phi}}
\right).
\end{equation}
Defining the effective potential through
\begin{equation}
\dot r^{\,2}+V_{\textrm eff}(r)=0,
\end{equation}
we obtain
\begin{equation}
V_{\textrm eff}(r)
=
-\frac{1}{g_{rr}(r)}
\left[
\frac{\left(\mathcal E-\frac{q}{\mu}\xi_t(r)\right)^2}{-g_{tt}(r)}
-
\left(
\epsilon
+
\frac{\mathcal L^2}{g_{\phi\phi}(r)}
\right)
\right].
\end{equation}

For a \emph{neutral} particle ($q=0$) the effective potential reduces to
\begin{equation}
V_{\textrm eff}(r)
=
-\frac{1}{g_{rr}(r)}
\left[
\frac{\mathcal E^2}{-g_{tt}(r)}
-
\left(
\epsilon
+
\frac{\mathcal L^2}{g_{\phi\phi}(r)}
\right)
\right].
\end{equation}
Substituting the Harrison scalings given by Eq.~\eqref{har}  we get,
\begin{equation}
V_{\textrm eff}(r)
=
-\frac{1}{\digamma^2(r,\theta) g^{(0)}_{rr}}
\left[
\frac{\mathcal E^2\digamma^2(r,\theta)}{A^2}
-
\epsilon
-
\frac{\mathcal L^2}{\digamma^2(r,\theta) g^{(0)}_{\phi\phi}}
\right].
\end{equation}
Therefore, in general $V_{\textrm eff}(r)\neq V^{(0)}_{\textrm eff}(r)$
whenever $\digamma(r,\theta)(r)\neq 1$.
However, since $A^2(r_h)=0$ implies $\digamma(r,\theta)(r_h)=1$,
the horizon position and horizon-based quantities coincide
with the uncharged case.

Circular equatorial orbits satisfy
\begin{equation}
V_{\textrm eff}(r_c)=0,
\qquad
\left.\frac{dV_{\textrm eff}}{dr}\right|_{r=r_c}=0,
\end{equation}
while the innermost stable circular orbit (ISCO) additionally obeys
\begin{equation}
\left.\frac{d^2V_{\textrm eff}}{dr^2}\right|_{r=r_{\textrm ISCO}}=0.
\end{equation}

The electromagnetic interaction enters exclusively through the
combination $\mathcal E-\frac{q}{\mu}\xi_t$.
Hence the geometry itself remains static and non-rotating,
and neutral null geodesics are governed purely by the metric.
In contrast, charged particle motion acquires a genuine
Coulomb-type contribution, modifying the orbital energy balance
and stability structure relative to the vacuum seed.
\begin{figure}
\centering
\subfigure[~ISCO radius as a function of the distortion parameter $B$.]{\label{fig:isco}\includegraphics[scale=0.21]{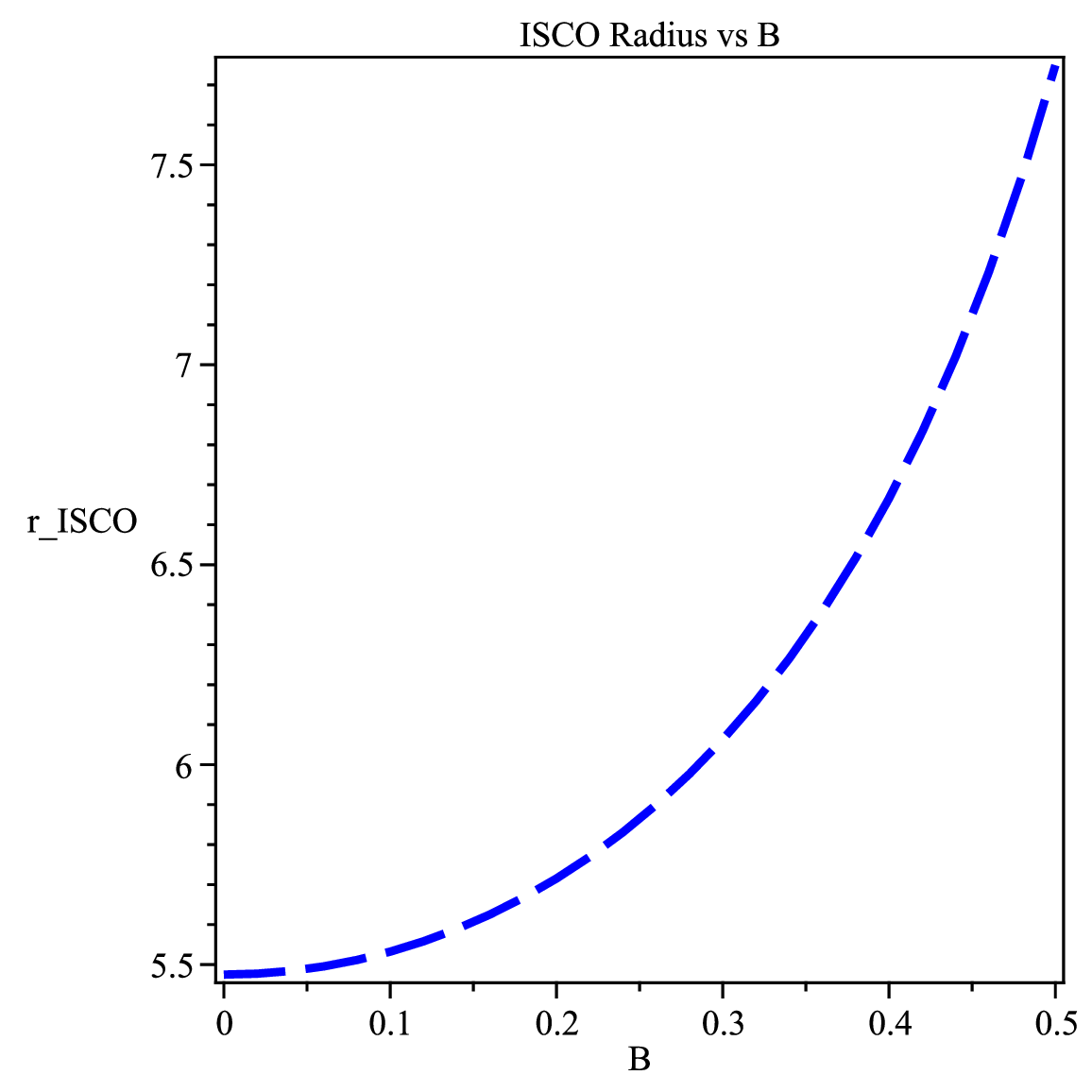}}
\subfigure[~Effective potential $V_{\textrm eff}(r)$ for different values of $B$.]{\label{fig:pot}\includegraphics[scale=0.21]{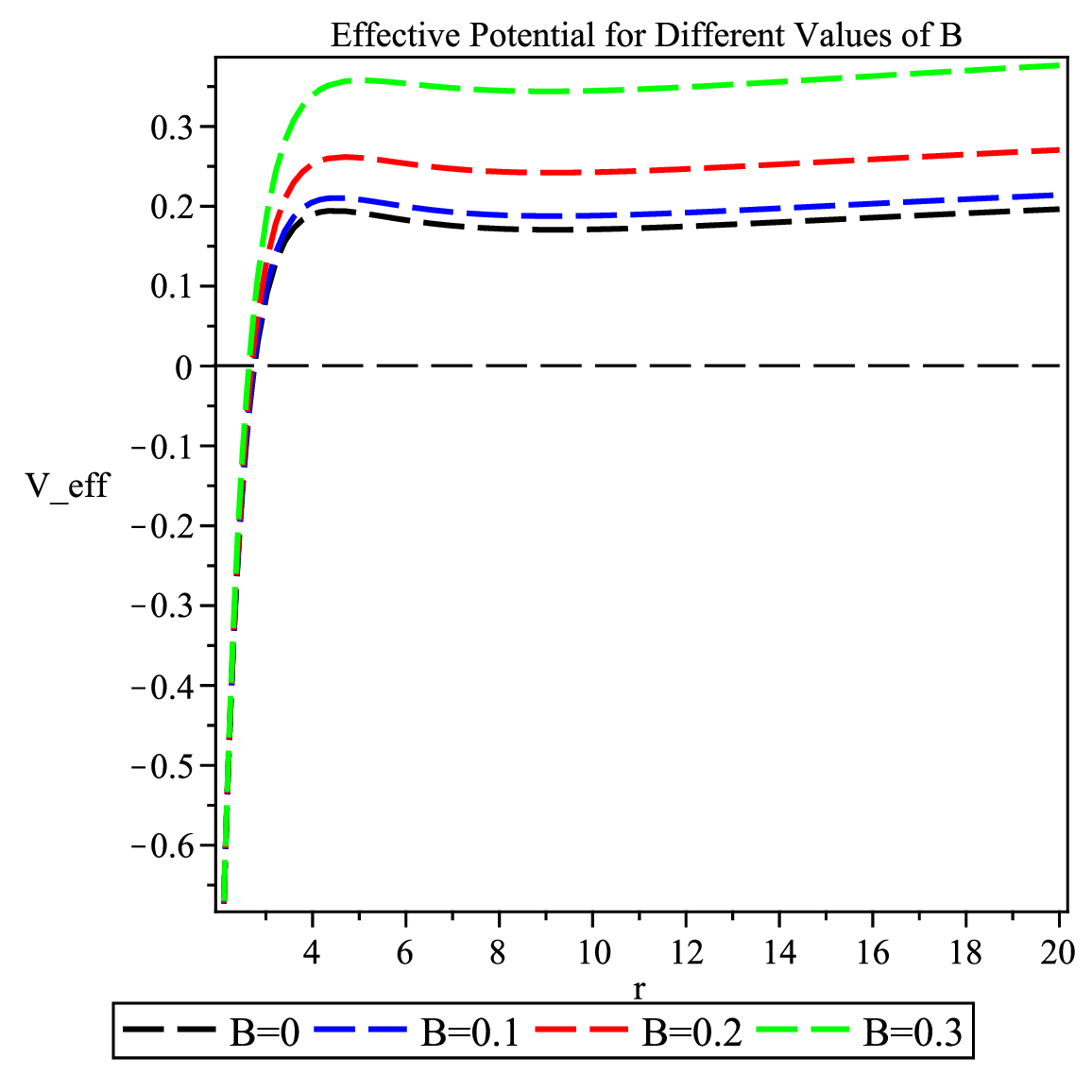}}
\subfigure[~First radial derivative of the effective potential $dV_{\textrm eff}/dr$.]{\label{fig:dpot}\includegraphics[scale=0.21]{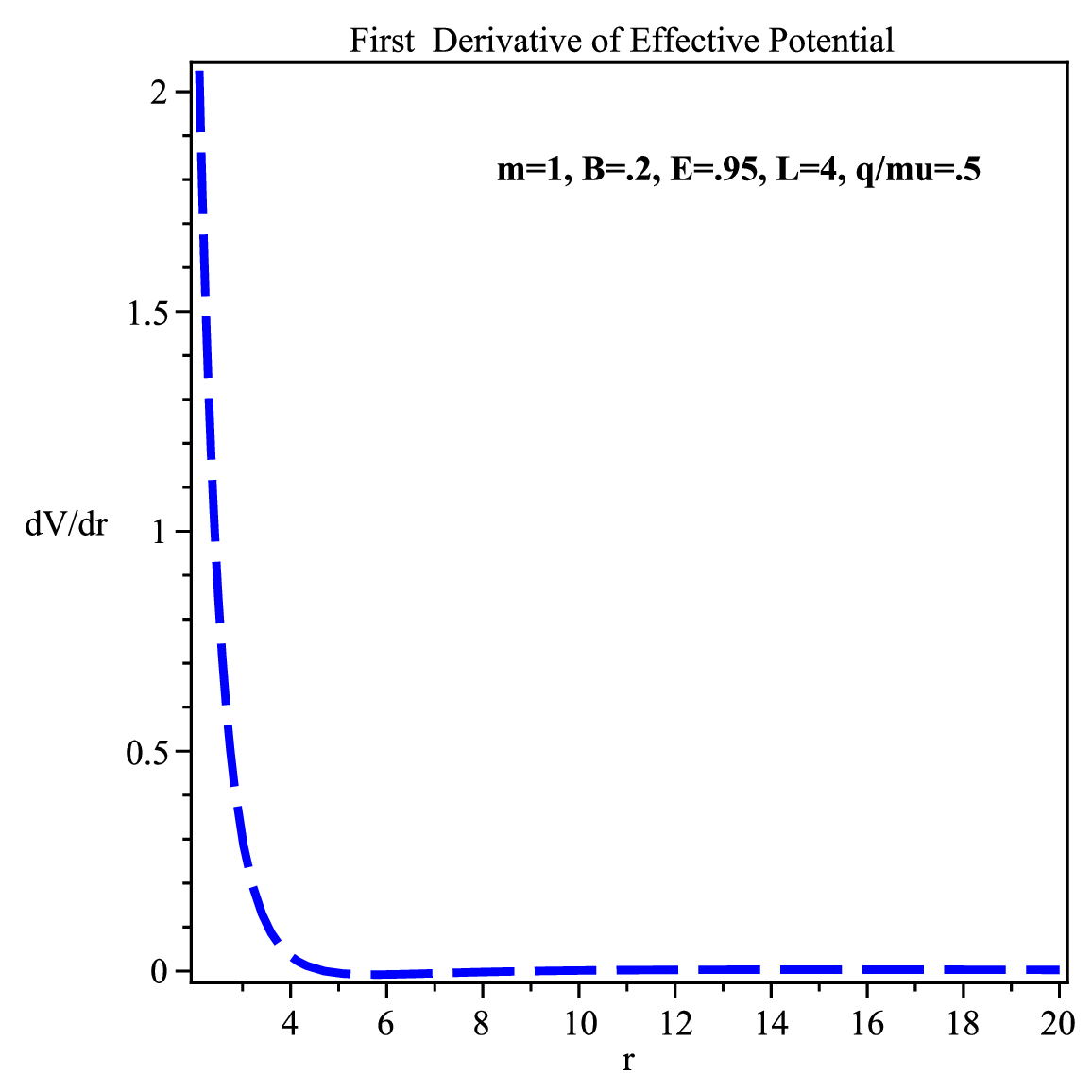}}
\subfigure[~Second radial derivative of the effective potential $d^2V_{\textrm eff}/dr^2$.]{\label{fig:ddpot}\includegraphics[scale=0.21]{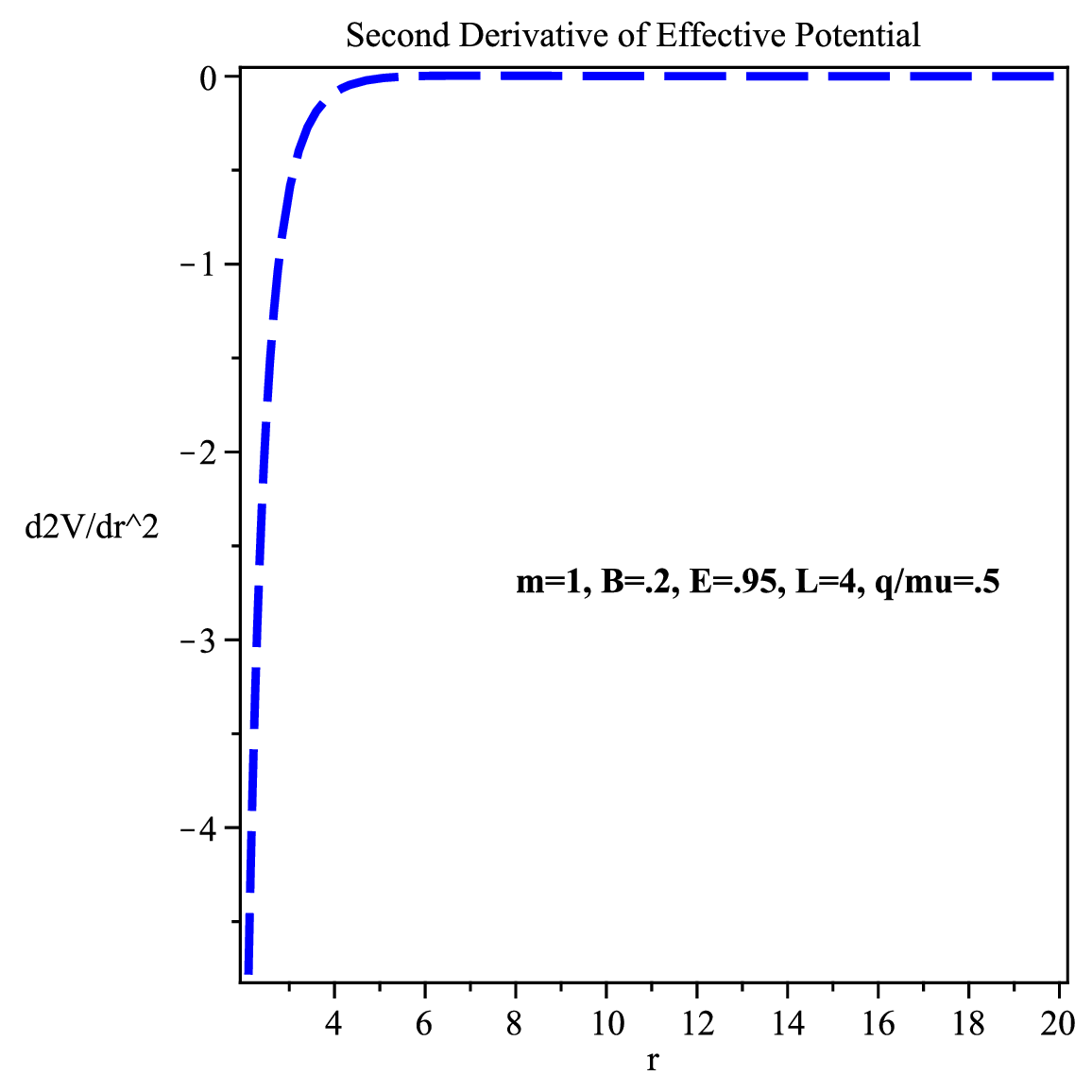}}
%
\caption{Illustration of particle dynamics and scalar perturbations in the charged distorted black hole spacetime.
Panel \subref{fig:isco} shows the dependence of the innermost stable circular orbit (ISCO) radius on the distortion
parameter $B$, demonstrating that the location of the marginally stable orbit shifts as the external
distortion increases.
Panel \subref{fig:pot} displays the effective potential for different values of $B$, indicating that the deformation
parameter modifies the depth and position of the potential well that governs particle motion.
Panels \subref{fig:dpot} and \subref{fig:ddpot} present the first and second derivatives of the effective potential, which determine
the conditions for circular orbits and their stability.
In particular, the condition $dV_{\textrm eff}/dr=0$ identifies circular trajectories, while
$d^2V_{\textrm eff}/dr^2>0$ signals stable orbits and $d^2V_{\textrm eff}/dr^2<0$ corresponds to unstable ones.}
\label{Fig:1}
\end{figure}

Figure~\ref{Fig:1} summarizes several dynamical aspects of the charged distorted
black hole spacetime. The behavior of the ISCO radius shown in panel
\subref{fig:isco} demonstrates that the external distortion parameter $B$
significantly affects the location of the innermost stable circular orbit.
As $B$ increases, the anisotropic deformation of the spacetime geometry
alters the gravitational potential experienced by test particles, leading
to a shift of the marginally stable orbit. This indicates that the presence
of external distortions modifies the structure of particle motion in the
near-horizon region.

The effective potential displayed in panel \subref{fig:pot} shows the
existence of a potential well outside the horizon. Such a well allows the
presence of bound particle trajectories and determines the radial region
where stable motion is possible. The depth and position of the potential
well depend on the parameters of the spacetime, particularly the distortion
parameter $B$ and the electromagnetic interaction.

The first and second derivatives of the effective potential, shown in
panels \subref{fig:dpot} and \subref{fig:ddpot}, provide the mathematical
conditions for circular motion and orbital stability \cite{Cardoso:2008bp}. A circular orbit
occurs at radii where the radial force vanishes, which corresponds to the
condition $dV_{\textrm eff}/dr = 0$. The stability of such an orbit is then
determined by the curvature of the potential: if
$d^2V_{\textrm eff}/dr^2 > 0$, the orbit corresponds to a local minimum of the
potential and is therefore stable, whereas if
$d^2V_{\textrm eff}/dr^2 < 0$ the orbit is unstable. The transition between
stable and unstable circular orbits defines the ISCO radius, which marks
the inner boundary of stable particle motion around the black hole.
\section{Black Hole Shadow}\label{VI}

In this section we investigate the shadow cast by the charged distorted black hole.
Since the spacetime is not asymptotically flat due to the presence of the external
distortion parameter $B$, the shadow is most naturally defined for a static observer
located at a finite radial position $r_{\textrm obs}$.
We restrict the analysis to the equatorial plane $\theta=\pi/2$, where the metric
functions simplify and the null geodesic equations admit a transparent interpretation.

For null geodesics, the Hamilton--Jacobi equation gives
\[
0 = g_{\mu\nu}\dot{x}^\mu \dot{x}^\nu ,
\]
together with the conserved quantities associated with stationarity and axisymmetry,
\[
E=-g_{tt}\dot{t}, \qquad L=g_{\phi\phi}\dot{\phi},
\]
where $E$ and $L$ are respectively the energy and angular momentum of the photon.
In the equatorial plane, the radial equation for null motion takes the form
\begin{align}
&g_{rr}\dot{r}^{\,2}
=
\frac{E^2}{-g_{tt}}
-
\frac{L^2}{g_{\phi\phi}}, \quad \mbox{by introducing the impact parameter}\quad
b \equiv \frac{L}{E},\quad \mbox{the radial equation becomes} \nonumber\\
& g_{rr}\dot{r}^{\,2}
=
E^2
\left(
\frac{1}{-g_{tt}}
-
\frac{b^2}{g_{\phi\phi}}
\right).
\end{align}
The boundary of the shadow is determined by unstable circular photon orbits,
which satisfy
\[
\dot{r}=0,
\qquad
\frac{d}{dr}\left(\dot{r}^{\,2}\right)=0 .
\]
From the first condition one obtains the critical impact parameter
\[
b_{\textrm ph}^2
=
\frac{g_{\phi\phi}(r_{\textrm ph})}{-g_{tt}(r_{\textrm ph})},
\]
where $r_{\textrm ph}$ denotes the radius of the photon orbit.
The second condition gives the photon-ring equation
\[
\frac{d}{dr}
\left(
\frac{g_{\phi\phi}}{-g_{tt}}
\right)_{r=r_{\textrm ph}}
=0 .
\]
Thus, the shadow is determined by the extrema of the ratio
$g_{\phi\phi}/(-g_{tt})$ evaluated in the equatorial plane.

For a static observer located at $(r_{\textrm obs},\theta=\pi/2)$, the angular radius
of the shadow can be obtained from the local orthonormal frame.
The celestial angle $\alpha_{\textrm sh}$ satisfies
\begin{align}
\sin^2\alpha_{\textrm sh}
=
\frac{b_{\textrm ph}^2\,[-g_{tt}(r_{\textrm obs})]}
     {g_{\phi\phi}(r_{\textrm obs})},
\quad \mbox{or equivalently}\quad
\sin^2\alpha_{\textrm sh}
=
\frac{g_{\phi\phi}(r_{\textrm ph})}{-g_{tt}(r_{\textrm ph})}
\,
\frac{-g_{tt}(r_{\textrm obs})}{g_{\phi\phi}(r_{\textrm obs})}.\label{shod1}
\end{align}
This expression shows explicitly how the shadow depends both on the location of
the photon orbit and on the position of the observer. Using the charged metric (\ref{char}) restricted to $\theta=\pi/2$, one has
\begin{align}
g_{tt}(r,\pi/2) = -\frac{\Sigma(r,\pi/2)\,f(r)}{\digamma(r,\pi/2)^2},
\qquad
g_{\phi\phi}(r,\pi/2)
=
\frac{4r^2 \digamma(r,\pi/2)^2 \Delta_\phi^2}
     {\left[1+\Omega_1(r,\pi/2)\right]^2},
\end{align}
where the functions $\Sigma(r,\theta)$, $f(r)$, $\digamma(r,\theta)$, and $\Omega_1(r,\theta)$
are defined in Sec.~\ref{II}.
Therefore the photon-ring radius is determined by
\[
\frac{d}{dr}
\left[
\frac{4r^2 \digamma(r,\pi/2)^4 \Delta_\phi^2}
{\Sigma(r,\pi/2)\,f(r)\,\left(1+\Omega_1(r,\pi/2)\right)^2}
\right]_{r=r_{\textrm ph}}
=0 .
\]
The corresponding critical impact parameter is
\[
b_{\textrm ph}^2
=
\frac{4r_{\textrm ph}^2 \digamma(r_{\textrm ph},\pi/2)^4 \Delta_\phi^2}
{\Sigma(r_{\textrm ph},\pi/2)\,f(r_{\textrm ph})\,
\left(1+\Omega_1(r_{\textrm ph},\pi/2)\right)^2}.
\]

These expressions show that both the distortion parameter $B$ and the electric
charge $Q$ modify the shadow through the metric functions $\Sigma$ and $z$.
In particular, the distortion parameter changes the location of the photon ring
through its effect on the angular sector of the geometry, while the electric field
contributes through the Harrison factor $z(r,\theta)$.
Consequently, the black hole shadow provides an observable probe of the interplay
between geometric distortion and electromagnetic charging in the present spacetime.

For the numerical analysis, one may fix a static observer at a finite radius
$r_{\textrm obs}$ and study the dependence of $\alpha_{\textrm sh}$ on the parameters
$B$ and $Q$. In this way one can determine how the external distortion and the
electric charge alter the apparent size of the shadow relative to the corresponding
vacuum or spherically symmetric limits.
\begin{figure}
\centering
\subfigure[~Black hole shadow for different values of the distortion parameter $B$]{\label{fig:shad}\includegraphics[scale=0.27]{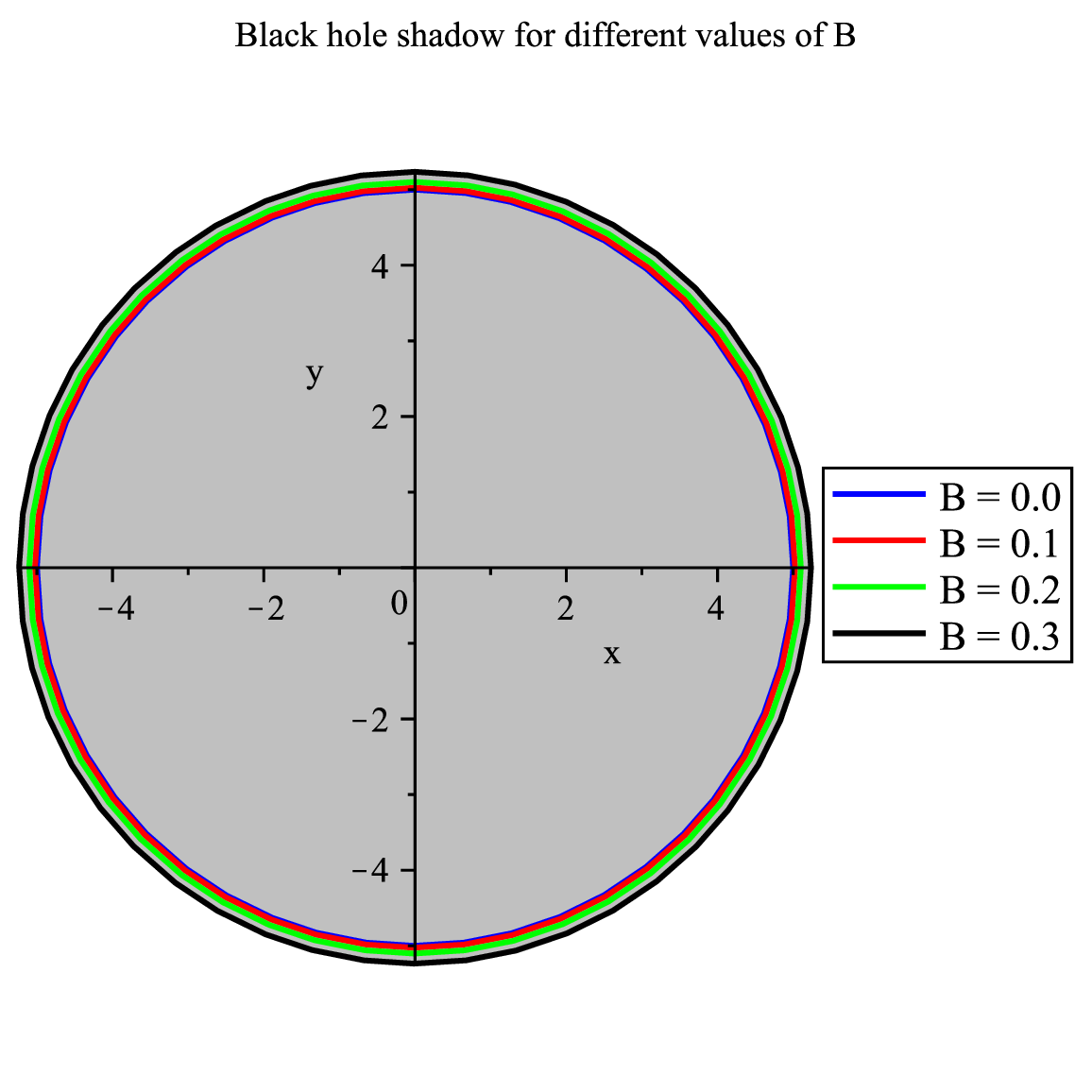}}
\subfigure[~Shadow angular radius $\alpha_{\textrm sh}(B)$ and $\alpha_{\textrm sh}(Q)$]{\label{fig:phsp}\includegraphics[scale=0.27]{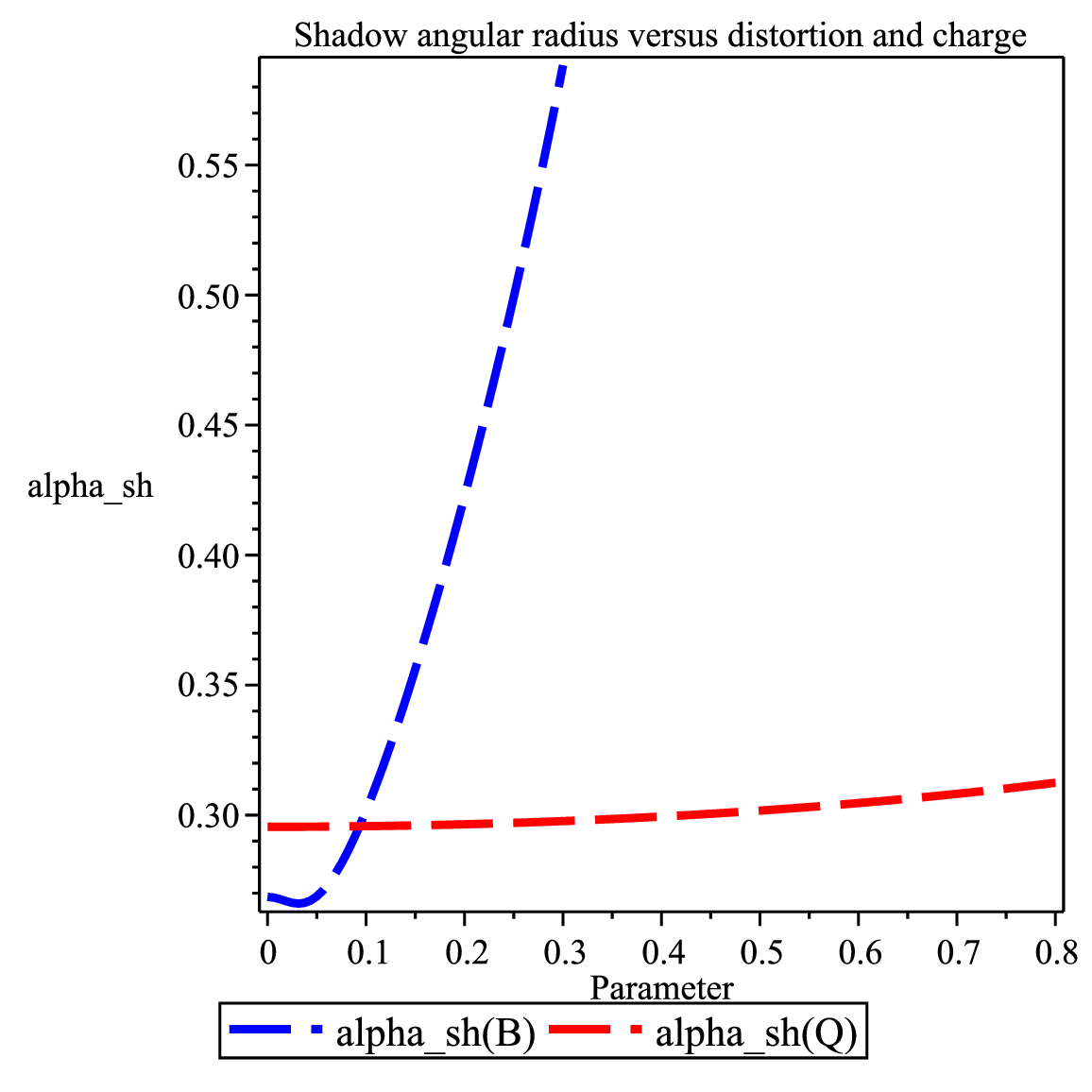}}
\caption{
Shadow properties of the charged distorted black hole.
Panel \subref{fig:shad} shows the shadow contour in the observer sky for several values of the distortion parameter $B$.
The curves correspond to $B=0.0$, $B=0.1$, $B=0.2$, and $B=0.3$.
As the distortion parameter increases, the shadow radius becomes slightly larger while preserving an approximately circular shape, indicating that the external distortion modifies the null geodesics that define the shadow boundary. Panel \subref{fig:phsp} displays the angular radius of the shadow $\alpha_{\textrm sh}$ as a function of the distortion parameter $B$ and the electric charge $Q$.
The blue dashed curve represents $\alpha_{\textrm sh}(B)$ for fixed charge $Q=0.5$, while the red dashed curve represents $\alpha_{\textrm sh}(Q)$ for fixed distortion parameter $B=0.1$. These curves are obtained from the finite-distance shadow formula given by Eq.\eqref{shod1}
where the photon sphere radius $r_{\textrm ph}$ is determined by the photon-ring condition
\(
\frac{d}{dr}\left(\frac{g_{\phi\phi}}{-g_{tt}}\right)=0.
\)
The blue curve shows that the shadow size increases with the distortion parameter, reflecting the outward displacement of the unstable photon orbit.
The red curve shows the corresponding variation with the electric charge, illustrating that the charge modifies the shadow through the Harrison factor in the metric.
The parameters used in the numerical evaluation are $m=1$, $\Delta_\phi=1$, and $r_{\textrm obs}=20$.
}
\label{Fig:3}
\end{figure}

Figure \ref{Fig:3} illustrates how the external distortion parameter affects
the optical appearance of the charged black hole. Panel \subref{fig:shad} shows
that the shadow remains nearly circular but its size increases
slightly as the distortion parameter $B$ grows. This behavior is
consistent with the outward shift of the unstable photon orbit,
which defines the boundary of the shadow.

Panel \subref{fig:phsp} provides a quantitative description of how the apparent shadow size depends on both the distortion parameter $B$ and the electric charge $Q$.
The blue dashed curve corresponds to the dependence $\alpha_{\textrm sh}(B)$ at fixed charge $Q=0.5$, whereas the red dashed curve represents $\alpha_{\textrm sh}(Q)$ at fixed distortion parameter $B=0.1$.
In both cases the angular shadow radius is computed from the finite-distance observer formula, with the photon sphere radius determined by the unstable circular null geodesics.

The increase of $\alpha_{\textrm sh}$ with $B$ indicates that the external distortion shifts the photon sphere outward and therefore enlarges the apparent shadow.
The dependence on $Q$ shows that the electric charge also affects the optical appearance of the black hole through its contribution to the Harrison factor $\digamma(r,\theta)$ appearing in the metric coefficients.
Thus, panel \subref{fig:phsp} demonstrates that both the geometric distortion and the electromagnetic charging leave observable imprints on the angular size of the black hole shadow.
These results demonstrate that external distortions leave a
measurable imprint on observable optical features such as the
black hole shadow.

\subsection{The geometric relation between the shadow and eikonal quasinormal modes}

An important consequence of the unstable circular photon orbit is that
it controls both the shadow boundary and the eikonal quasinormal-mode
spectrum. In the eikonal regime, the real part of the quasinormal
frequency is approximately determined by the angular frequency of the
photon orbit,
\begin{equation}
\omega_{\textrm QNM} \approx \ell \,\Omega_{\textrm ph}
- i\left(n+\frac12\right)\lambda ,
\end{equation}
where $\Omega_{\textrm ph}$ and $\lambda$ are respectively the orbital
frequency and the Lyapunov exponent of the unstable photon orbit.

For the present spacetime,
\begin{equation}
\Omega_{\textrm ph}
=
\sqrt{\frac{-g_{tt}(r_{\textrm ph})}{g_{\phi\phi}(r_{\textrm ph})}} .
\end{equation}
Using the finite-distance shadow formula derived in Eq.~(85),
\begin{equation}
\sin^2\alpha_{\textrm sh}
=
\frac{g_{\phi\phi}(r_{\textrm ph})}{-g_{tt}(r_{\textrm ph})}
\frac{-g_{tt}(r_{\textrm obs})}{g_{\phi\phi}(r_{\textrm obs})},
\end{equation}
one finds
\begin{equation}
\Omega_{\textrm ph}
=
\sqrt{\frac{-g_{tt}(r_{\textrm obs})}{g_{\phi\phi}(r_{\textrm obs})}}
\;\frac{1}{\sin\alpha_{\textrm sh}} .
\end{equation}
Therefore the real part of the eikonal quasinormal frequency satisfies
\begin{equation}
\frac{{\textrm Re}(\omega_{\textrm QNM})}{\ell}
\approx
\sqrt{\frac{-g_{tt}(r_{\textrm obs})}{g_{\phi\phi}(r_{\textrm obs})}}
\;\frac{1}{\sin\alpha_{\textrm sh}} .
\end{equation}
This relation shows explicitly that the same unstable photon orbit that
determines the shadow also governs the leading oscillation frequency of
the perturbative response.

\section{Superradiant Scattering of Charged Scalar Fields}\label{VII}

An important aspect of wave dynamics in black hole spacetimes is the
possibility of superradiant scattering. In such processes, incident
bosonic waves can extract energy from the black hole and emerge with an
amplified amplitude. Superradiance has been extensively investigated in
rotating and charged black holes, where it plays a key role in several
phenomena including wave amplification, black hole bombs, and possible
instabilities in the presence of confining mechanisms
\cite{Bekenstein:1973mi,Starobinskii:1973hgd,Brito:2015oca}.

For a charged scalar field with charge $q_s$ and frequency $\omega$
propagating in a charged black hole background, the superradiant
condition is generally expressed as
\begin{equation}
0 < \omega < q_s \Phi_H ,
\end{equation}
where $\Phi_H$ denotes the electric potential evaluated on the event
horizon \cite{Bekenstein:1973mi,Brito:2015oca}. When this inequality is
satisfied, the reflected wave carries more energy than the incident
wave, leading to an extraction of electromagnetic energy from the black
hole.

In the spacetime constructed in this work, the electromagnetic field is
generated through a Harrison transformation applied to a distorted
vacuum geometry. The resulting electromagnetic four-potential has only
a temporal component,
\begin{equation}
\xi_\mu dx^\mu = \xi_t(r,\theta)\, dt ,
\end{equation}
with
\begin{equation}
\xi_t(r,\theta) = \frac{E_0 A^2(r,\theta)}{z(r,\theta)} .
\end{equation}
Here $A^2(r,\theta)=\Sigma(r,\theta)f(r)$ represents the squared lapse
function of the seed metric, while $z(r,\theta)$ denotes the nonlinear
factor introduced by the Harrison transformation within the
Einstein-Maxwell solution-generating framework
\cite{Harrison:1968wue,Stephani:2003tm}.

The event horizon of the spacetime is located at the radius where
$f(r_h)=0$. Since the function $A^2(r,\theta)$ is proportional to
$f(r)$, it follows that
\begin{equation}
A^2(r_h,\theta)=0 .
\end{equation}
Substituting this result into the expression for the electromagnetic
potential immediately yields
\begin{equation}
\xi_t(r_h,\theta)=0 .
\end{equation}
Therefore the electric potential evaluated on the horizon vanishes,
\begin{equation}
\Phi_H = \xi_t(r_h,\theta)=0 .
\end{equation}

As a consequence, the superradiant condition becomes
\begin{equation}
0 < \omega < 0 ,
\end{equation}
which cannot be satisfied for any real frequency. We therefore conclude
that charged scalar waves propagating in the present
Harrison charged distorted black hole spacetime do not exhibit
superradiant amplification.

This result reflects a distinctive feature of the Harrison charging
procedure applied to distorted geometries. Although the transformation
introduces a nontrivial electromagnetic field in the exterior region,
the gauge potential is proportional to the lapse function of the seed
metric and therefore vanishes on the horizon. As a consequence, the
mechanism responsible for charged superradiant amplification in
Reissner-Nordstr\"om black holes is absent in the present configuration.
The charged distorted black hole solution studied here is therefore
stable against superradiant extraction of energy by charged scalar
perturbations.

\section{Charged scalar perturbations}\label{VIII}
We consider a complex scalar field $\Psi$ of mass $\mu_s$ and charge $q_s$
propagating on the charged background. The field satisfies the minimally
coupled Klein--Gordon equation
\begin{equation}
\frac{1}{\sqrt{-g}}
D_\mu\!\left(\sqrt{-g}\,g^{\mu\nu}D_\nu\Psi\right)
- \mu_s^2 \Psi = 0,
\qquad
D_\mu := \nabla_\mu - i q_s \xi_\mu .
\end{equation}
Since the gauge field has only a temporal component,
\(
\xi_\mu dx^\mu = \xi_t(r,\theta)\,dt,
\)
the covariant derivatives reduce to
\(
D_t = \partial_t - i q_s \xi_t(r,\theta),
\qquad
D_\phi = \partial_\phi \).
Using stationarity and axisymmetry, we decompose
\begin{equation}
\Psi(t,r,\theta,\phi)
=
e^{-i\omega t}\, e^{i m \phi}\, \psi(r,\theta),
\qquad m\in\mathbb{Z}, \quad\mbox{then} \quad
D_t \Psi = -i(\omega - q_s \xi_t)\Psi,
\qquad
D_\phi \Psi = i m \Psi.
\end{equation}
Substituting into the field equation yields
\begin{align}
\frac{1}{\sqrt{-g}}
\partial_r\!\left(\sqrt{-g}\,g^{rr}\partial_r\psi\right)
+
\frac{1}{\sqrt{-g}}
\partial_\theta\!\left(\sqrt{-g}\,g^{\theta\theta}\partial_\theta\psi\right)
+
\Big[
g^{tt}(\omega - q_s \xi_t)^2
+
\frac{m^2}{g_{\phi\phi}}
-
\mu_s^2
\Big]\psi
= 0.
\end{align}
Since the background is not separable in $r$ and $\theta$, we restrict the analysis to equatorial perturbations at $\theta=\pi/2$, assuming
\(
\left.\partial_{\theta}\psi\right|_{\theta=\pi/2}=0 \).
Defining
\(
R(r)\equiv \psi(r,\pi/2)
\)
the wave equation reduces to
\begin{equation}
\frac{1}{\sqrt{-g}}
\frac{d}{dr}\!\left(\sqrt{-g}\,g^{rr}\frac{dR}{dr}\right)
+
\left[
g^{tt}(\omega - q_s \xi_t(r))^2
+
\frac{m^2}{g_{\phi\phi}(r)}
-
\mu_s^2
\right]R
=0.
\end{equation}
Defining
\[
R(r)=\frac{u(r)}{\sqrt{\mathcal K(r)}},
\qquad
\mathcal K(r)=\sqrt{-g}\,g^{rr},
\]
the radial equation becomes
\begin{equation}
\frac{d^2u}{dr^2}
+
\left[
Q(r)\,(\omega - q_s \xi_t(r))^2
+
U_{\textrm geo}(r)
\right]u
=0,\quad \mbox{where}\quad
Q(r)= -\frac{g^{tt}}{g^{rr}},
\end{equation}
and the purely geometric contribution is
\[
U_{\textrm geo}(r)
=
\frac{1}{g^{rr}}
\left(
\frac{m^2}{g_{\phi\phi}} - \mu_s^2
\right)
-
\frac{1}{2}\frac{\mathcal K''}{\mathcal K}
+
\frac{1}{4}\left(\frac{\mathcal K'}{\mathcal K}\right)^2.
\]
 Introducing the tortoise coordinate
\begin{equation}
\frac{dr_*}{dr}=\sqrt{\frac{g_{rr}}{-g_{tt}}},\quad \mbox{
the radial equation takes the Schr\"odinger-like form} \quad
\frac{d^2u}{dr_*^2}
+
\left[
\bigl(\omega-q_s\xi_t(r)\bigr)^2
-
V_{\mathrm{sc}}(r)
\right]u=0.\label{sch}
\end{equation}
The weak-coupling approximation is justified when the electromagnetic interaction term $q_s \xi_t$ remains small compared to the characteristic oscillation frequency near the peak of the effective potential barrier. In this regime the dominant structure of the wave equation is governed by the geometric potential $V_{sc}(r)$, while the gauge interaction introduces only a perturbative correction to the frequency. Consequently, the standard WKB formalism can be consistently applied to estimate the quasinormal spectrum.

For a charged scalar field, the radial master equation takes the form given by Eq.~\eqref{sch} which in general, because $\xi_t(r)$ depends on $r$, the wave equation is not strictly of the standard WKB form with a purely frequency-independent potential. In this work, we therefore adopt the weak-coupling approximation for the scalar charge, assuming that the radial variation of the gauge term is subleading near the peak of the barrier. Under this approximation, the dominant potential barrier is provided by $V_{\textrm sc}(r)$, while the electromagnetic coupling enters as a perturbative correction through the gauge-invariant combination $(\omega-q_s\xi_t)$.

Therefore, in the quasinormal-mode analysis, we work directly with Eq.~(\ref{sch}) rather than introducing an explicitly frequency-dependent effective potential.

The influence of the background electric field enters entirely
through the gauge-invariant combination
$\omega - q_s \xi_t(r)$. For single-barrier potentials, the quasinormal modes can then be estimated using the WKB approximation. In the weak-coupling regime, the peak of the barrier is determined from the frequency-independent part of the potential,
\[
\frac{dV_{\textrm sc}}{dr_*}\bigg|_{r_{*0}}=0.
\]
We define
\[
V_0 := V_{\textrm sc}(r_{*0}),
\qquad
V_0'' := \frac{d^2V_{\textrm sc}}{dr_*^2}\bigg|_{r_{*0}}.
\]
At leading order one finds
\[
\omega^2 \approx V_0 - i\left(n+\tfrac12\right)\sqrt{-2V_0''},
\qquad n=0,1,2,\dots
\]
For improved accuracy one may use the third-order WKB correction,
\[
\omega^2
=
V_0
-
i\left(n+\tfrac12\right)\sqrt{-2V_0''}
+
\digamma_2
+
\digamma_3,
\]
where $\digamma_2$ and $\digamma_3$ are the second- and third-order WKB correction terms depending on the higher derivatives
$V_0^{(3)}, V_0^{(4)}, V_0^{(5)}$, and $V_0^{(6)}$ evaluated at the peak.

For numerical evaluation we consider:
\[
\mu_s=0,
\qquad
\ell=1,2,
\qquad
m=0,
\]
with background parameters
\[
m=1,
\qquad
B=0.1,
\qquad
Q=0.5,
\qquad
\Delta_\phi=1,
\]
and scalar charges
\[
q_s = -0.6,\; 0,\; 0.6.
\]

\begin{figure}
\centering
\subfigure[~ Scalar effective potential $V_{sc}(r)$]{\label{fig:11}\includegraphics[scale=0.21]{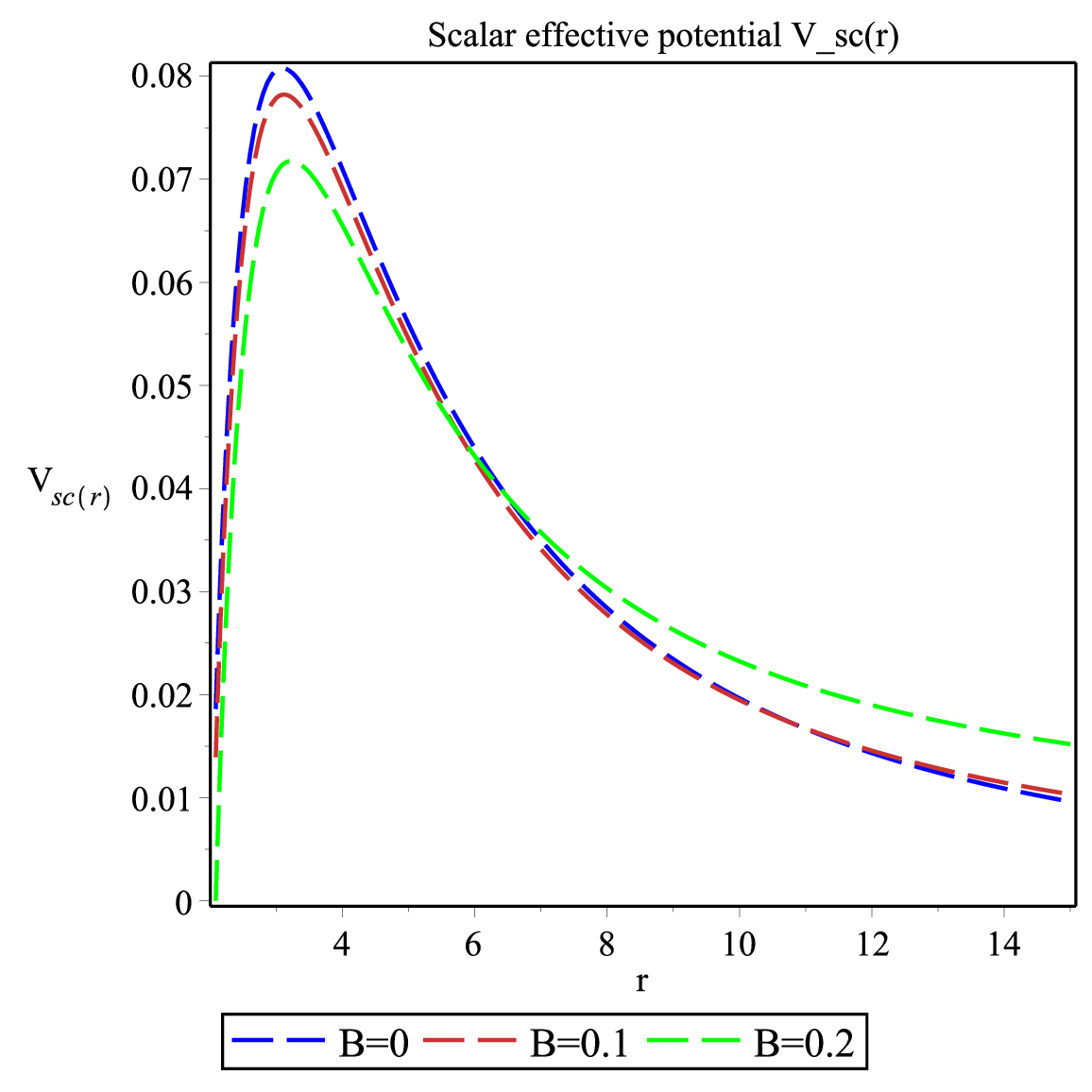}}
\subfigure[~Gauge-induced frequency shift $(\omega-q_s\xi_t)$]{\label{fig:22}\includegraphics[scale=0.21]{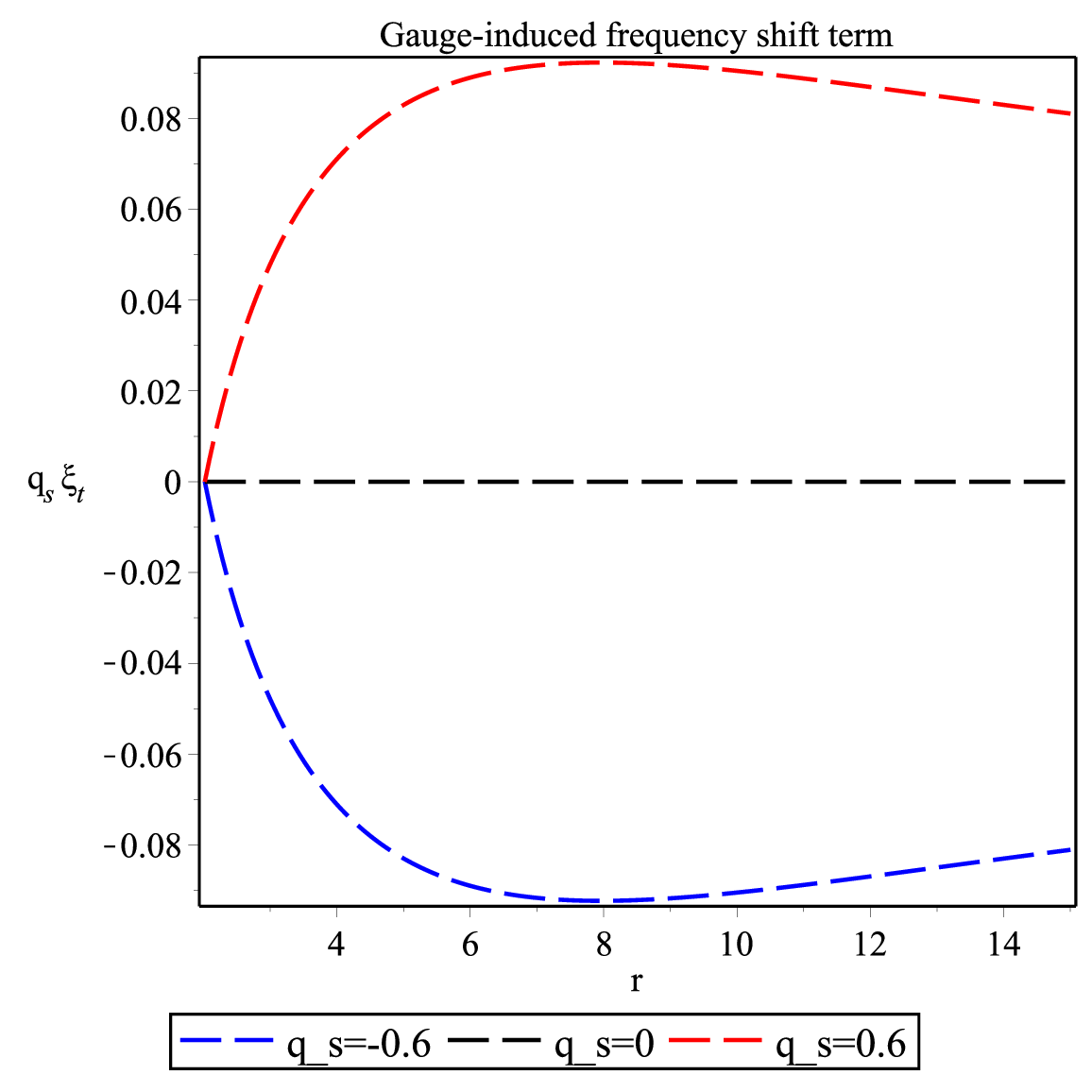}}
\subfigure[~Approximate QNM frequency versus scalar charge $q_s$]{\label{fig:33}\includegraphics[scale=0.21]{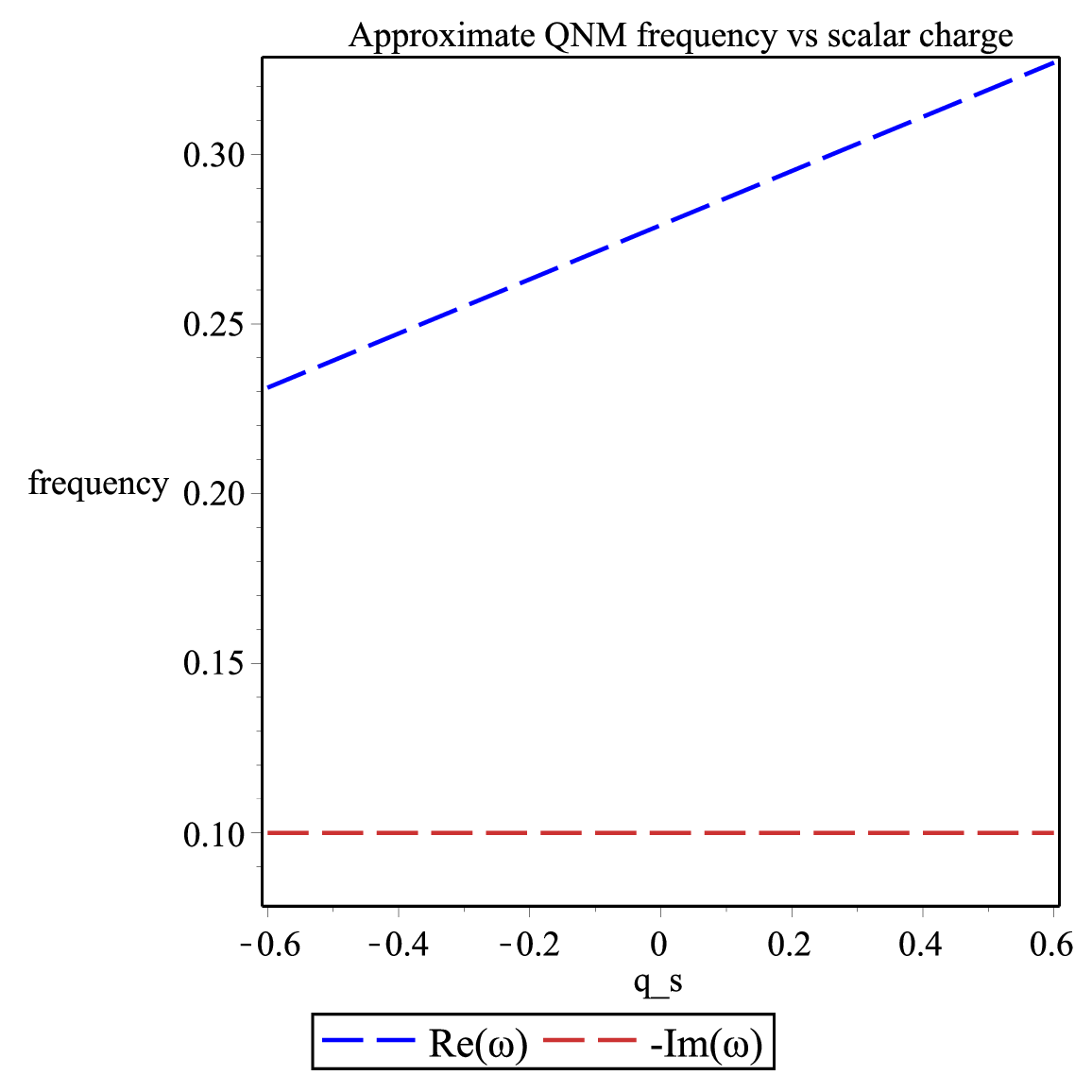}}
\subfigure[~Charged scalar barrier proxy]{\label{fig:44}\includegraphics[scale=0.21]{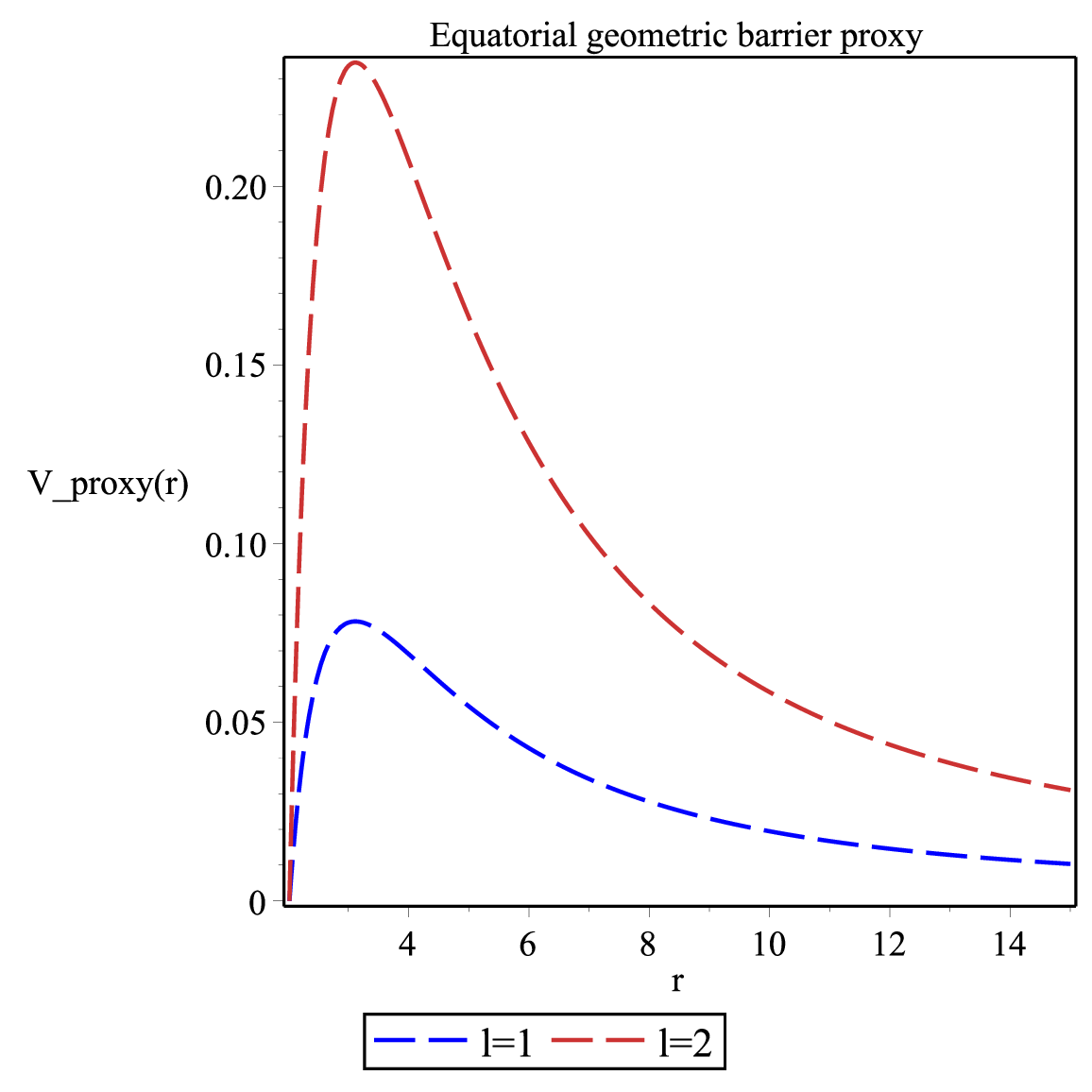}}
\caption{
Illustration of scalar perturbations in the charged distorted black hole spacetime.
\subref{fig:11} Scalar effective potential $V_{sc}(r)$ evaluated on the equatorial plane for several values of the scalar charge $q_s$.
The potential exhibits a single-barrier structure outside the event horizon, which is the typical condition required for the WKB approximation used in quasinormal-mode calculations.
\subref{fig:22} Gauge-induced frequency shift $(\omega - q_s \xi_t)$ showing the effect of the electromagnetic interaction between the charged scalar field and the background electric potential.
The coupling modifies the effective oscillation frequency experienced by the perturbation.
\subref{fig:33} Approximate quasinormal-mode frequency as a function of the scalar charge $q_s$ obtained within the weak-coupling WKB approximation.
The real part of the frequency varies with $q_s$, indicating that the electromagnetic interaction shifts the oscillation frequency of the perturbation.
\subref{fig:44} Illustration of the effective potential barrier governing the radial wave equation.
The presence of a single smooth peak confirms that the radial equation admits the standard WKB treatment for estimating the quasinormal spectrum.
}
\label{Fig:2}
\end{figure}
Although the horizon radius and temperature coincide with the
uncharged seed solution,
the wave dynamics does not.
The electric interaction modifies the effective barrier
and consequently shifts both the oscillation frequency
$\mathrm{Re}(\omega)$
and the damping rate
$-\mathrm{Im}(\omega)$.
Therefore, the QNM spectrum provides a clear dynamical
signature of the charged deformation.

Figure~\ref{Fig:2} illustrates several aspects of the charged scalar perturbations in the distorted black hole background.
Panel \subref{fig:11} shows the effective scalar potential $V_{sc}(r)$ obtained from the Schr\"odinger-like radial equation.
The potential forms a single barrier outside the event horizon, which ensures that the quasinormal-mode problem is well posed and suitable for the WKB approximation.

Panel \subref{fig:22} highlights the role of the electromagnetic interaction through the gauge-invariant combination $(\omega - q_s \xi_t)$ appearing in the wave equation.
This term describes the coupling between the scalar charge and the background electric field generated by the Harrison transformation.

Panel \subref{fig:33} presents the resulting quasinormal frequencies as a function of the scalar charge.
Within the weak-coupling approximation adopted in this work, the electromagnetic interaction shifts the oscillation frequency of the perturbation while leaving the overall barrier structure essentially unchanged.

Finally, panel \subref{fig:44} shows the effective potential barrier that governs the radial propagation of the perturbation.
The presence of a smooth single-peak potential confirms that the system satisfies the standard conditions required for the application of the WKB method in quasinormal-mode calculations.

\begin{table}[h]
\centering
\caption{Fundamental quasinormal frequencies of neutral scalar perturbations
for different multipole numbers $\ell$ in the weak-coupling WKB approximation,
with $m=1$, $B=0.1$, $Q=0.5$, $\Delta_\phi=1$, $q_s=0$, and $n=0$.}
\begin{tabular}{ccc}
\hline
$\ell$ & $\mathrm{Re}(\omega)$ & $-\mathrm{Im}(\omega)$ \\
\hline
1 & 0.29677 & 0.09926 \\
2 & 0.49524 & 0.10302 \\
3 & 0.69292 & 0.10413 \\
\hline
\end{tabular}
\label{tab:l_modes}
\end{table}
Table~\ref{tab:l_modes} shows the dependence of the fundamental
quasinormal frequency on the multipole number $\ell$. The real part of
the frequency increases with $\ell$, reflecting the larger oscillation
rate of modes with higher angular momentum, while the damping rate
changes only mildly for fixed overtone number $n=0$. This behavior is
consistent with the expected eikonal scaling of black hole
perturbations and supports the geometric interpretation of the
quasinormal spectrum in terms of unstable photon orbits.

\newpage

\section{Conclusions and perspectives}\label{IX}
Before summarizing the main results, it is useful to compare the physical
properties of the charged distorted black hole obtained in this work
with representative configurations previously studied in the literature.
The main differences are summarized in Table~\ref{tab:comparison}.
\begin{table*}[t]
\centering
\caption{Comparison between representative studies of distorted or
charged black holes and the present work.}
\begin{tabular}{lcccccc}
\hline
Study & Distortion & Electric charge & Thermodynamics & Particle motion & QNMs & Superradiance \\
\hline

Geroch \& Hartle (1982) \cite{Geroch:1982bv}
& Yes & No & No & No & No & No \\

Ernst (1976) \cite{Ernst:1967by}
& No & Magnetic field & No & No & No & No \\

Astorino (2026) \cite{Astorino:2026okd}
& Yes & No & Yes & No & No & No \\

Typical Reissner--Nordstr\"om studies \cite{1918KNAB...20.1238N}
& No & Yes & Yes & Yes & Yes & Yes \\

\textbf{Present work}
& Yes & Yes & Yes & Yes & Yes & No \\

\hline
\end{tabular}
\label{tab:comparison}
\end{table*}
In this work we construct an electrically charged extension of a distorted black hole spacetime using a Harrison transformation within the Einstein-Maxwell framework. A notable feature of the resulting geometry is that the electromagnetic field modifies the exterior spacetime through a nonlinear conformal factor while leaving the radial structure of the seed metric unchanged. Consequently, the location of the Killing horizon remains determined solely by the seed geometry, in contrast with the Reissner-Nordstr\"om solution where the electric charge directly shifts the horizon radius.  \cite{Harrison:1968wue,Stephani:2003tm}. The resulting geometry therefore represents a charged deformation of the distorted vacuum black hole solution.

We analyzed several geometric and physical properties of the charged spacetime. In particular, we showed that the location of the Killing horizon remains determined solely by the radial function of the seed metric, implying that the electric charge introduced by the Harrison transformation does not shift the horizon radius. This behavior differs from the standard Reissner--Nordstr\"om black hole, where the charge directly modifies the radial structure of the horizon \cite{Reissner:1916cle,Nordstrom:2018acn}. The thermodynamic quantities of the solution were also derived explicitly. We found that the entropy and temperature depend only on the geometric parameters of the seed spacetime, while the electromagnetic sector contributes through the electric potential and modifies the thermodynamic phase structure. These features are consistent with the general framework of black hole thermodynamics developed in \cite{Bekenstein:1973ur,Hawking:1976de,Wald:1993nt}.

The dynamics of charged test particles were studied through the effective potential formalism. The distortion parameter alters the gravitational potential and therefore modifies the structure of circular orbits and the location of the innermost stable circular orbit. Such modifications of particle dynamics in nontrivial black hole geometries are important for understanding astrophysical processes occurring in distorted gravitational environments \cite{Poisson:2009pwt,Frolov:1998wf}.

An additional observable feature of the charged distorted black hole is
its optical appearance as characterized by the black hole shadow.
Our analysis shows that the external distortion parameter modifies the
structure of the unstable photon orbit and consequently affects the
apparent angular size of the shadow for a static observer located at
finite distance. In particular, the distortion parameter tends to shift
the photon sphere outward, leading to a modest enlargement of the
shadow radius. The electric charge also contributes through the
Harrison factor appearing in the metric functions, which alters the
null geodesics that determine the shadow boundary. Although the shadow
remains nearly circular for the parameter range considered, the combined
effects of distortion and electromagnetic charging leave measurable
imprints on the shadow size. These results suggest that optical
observables such as black hole shadows may provide useful probes of
distorted black hole geometries and their electromagnetic extensions.

Another noteworthy property of the Harrison-charged distorted black
hole concerns the absence of charged superradiant amplification.
Because the electromagnetic potential vanishes on the event horizon,
the usual superradiant condition cannot be satisfied. This feature
distinguishes the present geometry from the Reissner-Nordstr\"om
solution and indicates that the spacetime is stable against
superradiant energy extraction by charged scalar perturbations.

Finally, we investigated the propagation of a charged scalar field in this background and derived the corresponding radial master equation governing scalar perturbations. In the weak-coupling regime, the quasinormal-mode spectrum was estimated using the WKB approximation \cite{Iyer:1986np,Konoplya:2003ii}. The electromagnetic interaction enters through the gauge-invariant combination $(\omega - q_s \xi_t)$ and produces shifts in the oscillation frequencies of the perturbations. Since quasinormal modes encode characteristic information about the background geometry, these results illustrate how external distortions and electromagnetic interactions influence the dynamical response of black holes \cite{Kokkotas:1999bd,Berti:2009kk}.

Overall, the present analysis demonstrates that charged distorted black holes generated through Harrison transformations exhibit a rich interplay between geometry, thermodynamics, particle dynamics, and perturbative stability. Such configurations provide a useful theoretical framework for exploring how external fields and electromagnetic interactions modify the physical properties of black hole spacetimes

Future investigations may extend the present analysis in several directions.
First, a more detailed study of the quasinormal-mode spectrum could be performed using numerical methods such as continued-fraction techniques or time-domain evolution, allowing a precise determination of the frequency spectrum beyond the weak-coupling approximation \cite{Leaver:1985ax,Konoplya:2011qq}.
Second, it would be interesting to explore the behavior of other perturbative fields, including electromagnetic and gravitational perturbations, in order to investigate the full stability properties of the charged distorted spacetime.

Beyond the geometric construction, we investigate several physical properties of the charged distorted black hole. In particular, we analyze the thermodynamic behavior of the horizon, the motion of charged test particles and the structure of circular orbits, the optical appearance of the black hole shadow for observers at finite distance, and the propagation of charged scalar perturbations together with their quasinormal-mode spectrum. These analyses illustrate how the interplay between external gravitational distortions and electromagnetic interactions influences the geometric, thermodynamic, optical, and dynamical properties of black hole spacetimes \cite{Cunha:2018acu,Berti:2015itd}.

\section*{Acknowledgements}
SC  acknowledges the support of  Istituto Nazionale di Fisica Nucleare (INFN) Sez. di Napoli, Iniziative Specifiche QGSKY and MOONLIGHT2.
and   the Gruppo Nazionale di Fisica Matematica (GNFM) of Istituto Nazionale di Alta Matematica (INDAM).This paper is based upon work from COST Action CA21136 {\it Addressing observational tensions in cosmology with systematics  and fundamental physics} (CosmoVerse) supported by COST (European Cooperation in Science and Technology).

\appendix 

\section{ Small-$B$ Asymptotic Expansion of the Invariants}\label{X}

We consider the weak-distortion regime
\[
\varepsilon := B^2 \ll 1,
\qquad
c := \cos\theta,
\qquad
s := \sin\theta,
\qquad
f_0(r) := 1 - \frac{2m}{r}.
\]

\subsubsection*{Expansion of the metric functions}

The radial lapse expands as
\begin{equation}
f(r)
=
f_0(r)
+
\varepsilon \left( f_0(r)\, r^2 - m^2 \right)
+
\mathcal O(\varepsilon^2).
\end{equation}
The angular function becomes
\begin{equation}
\Omega_1(r,\theta)
=
1
+
\frac{\varepsilon}{2}
\left(
r^2 s^2 + 2 m r c^2
\right)
+
\mathcal O(\varepsilon^2), \quad \mbox{
The conformal factor is}\quad
\Sigma(r,\theta)
=
1
-
\varepsilon
\left(
\frac{3}{2} r^2 s^2 + 2 m r c^2
\right)
+
\mathcal O(\varepsilon^2).
\end{equation}

Hence the squared lapse
\begin{equation}
A^2(r,\theta)
=
\Sigma f
=
f_0(r)
+
\varepsilon
\left[
\frac{r^2}{2}(-1+3c^2) f_0(r)
-
m^2
\right]
+
\mathcal O(\varepsilon^2).
\end{equation}
Its derivatives are
\begin{align}
\partial_r A^2
&=
\frac{2m}{r^2}
+
\varepsilon (1-3c^2)(m-r)
+
\mathcal O(\varepsilon^2),\quad
\partial_\theta A^2=
\varepsilon\, 3r(2m-r) s c
+
\mathcal O(\varepsilon^2).
\end{align}

\subsubsection*{Charging data}

\begin{equation}
E_0
=
\frac{Q(1+B^2 m^2)}{2m}
=
\frac{Q}{2m}
+
\varepsilon \frac{Qm}{2}
+
\mathcal O(\varepsilon^2),\qquad
E_0^2
=
\frac{Q^2}{4m^2}
+
\varepsilon \frac{Q^2}{2}
+
\mathcal O(\varepsilon^2).
\end{equation}

Define the $B=0$ charging factor
\begin{equation}
\digamma(r,\theta)_0(r)
=
1
-
\frac{Q^2}{4m^2}
\left(
1 - \frac{2m}{r}
\right).
\end{equation}

Then
\begin{equation}
\digamma(r,\theta)
=
\digamma(r,\theta)_0(r)
+
\varepsilon \digamma(r,\theta)_2(r,\theta)
+
\mathcal O(\varepsilon^2),\quad \mbox{with} \quad
\digamma(r,\theta)_2(r,\theta)
=
-
\frac{Q^2}{4m^2}
\left[
\frac{r^2}{2}(-1+3c^2) f_0(r)
-
m^2
\right]
-
\frac{Q^2}{2} f_0(r).
\end{equation}

\subsubsection*{The Maxwell invariant}

To order $\mathcal O(B^2)$,
\begin{equation}
{
F_{\mu\nu}F^{\mu\nu}
=
-
\frac{2Q^2}{r^4 \digamma(r,\theta)_0(r)^4}
\left[
1
+
\varepsilon\, \mathcal C_F(r,\theta)
\right]
+
\mathcal O(\varepsilon^2)
}
\end{equation}

where
\begin{equation}
\mathcal C_F(r,\theta)
=
2m^2
+
\frac{r^2}{m}(1-3c^2)(m-r)
+
3r^2 s^2
+
4mr c^2
-
4\frac{\digamma(r,\theta)_2(r,\theta)}{\digamma(r,\theta)_0(r)}.
\end{equation}

\subsubsection*{The Ricci invariant}

\begin{equation}
{
R_{\mu\nu}R^{\mu\nu}
=
\kappa^2
\frac{Q^4}{r^8 \digamma(r,\theta)_0(r)^8}
\left[
1
+
\varepsilon\, \mathcal C_R(r,\theta)
\right]
+
\mathcal O(\varepsilon^2)
}
\end{equation}

with
\begin{equation}
\mathcal C_R(r,\theta)
=
4m^2
+
2\frac{r^2}{m}(1-3c^2)(m-r)
+
6r^2 s^2
+
8mr c^2
-
8\frac{\digamma(r,\theta)_2(r,\theta)}{\digamma(r,\theta)_0(r)}.
\end{equation}
\begin{equation}
K(r,\theta)
=
\frac{48 m^2}{r^6}
+
\frac{8 Q^4}{r^8 \digamma(r,\theta)_0(r)^8}
+
\varepsilon K_2(r,\theta)
+
\mathcal O(\varepsilon^2),
\end{equation}

with
\begin{equation}
K_2(r,\theta)
=
\frac{96 m^2}{r^4}(1-3\cos^2\theta)
+
\frac{32 Q^4}{r^6 \digamma(r,\theta)_0(r)^8}(1-3\cos^2\theta)
-
\frac{64 Q^4}{r^8 \digamma(r,\theta)_0(r)^9}\digamma(r,\theta)_2(r,\theta),
\end{equation}

\subsubsection*{Remaining invariants}

For the static purely electric configuration, we have
\begin{equation}
R = 0,
\qquad
{}^\star R_{\mu\nu\rho\sigma} R^{\mu\nu\rho\sigma} = 0.
\end{equation}
%

\end{document}